\documentclass{pas}

\usepackage{multirow}
\usepackage{aas_macros}
\usepackage{graphicx}

\newcommand{\sjfull}{Swift~J1727.8-1613}

\newcommand{\difmap}{\textsc{difmap}}
\newcommand{\aips}{\textsc{aips}}
\newcommand{\distanceNoErr}{\qty{5.5}{\kilo\parsec}}
\newcommand{\distanceBurridge}{$5.5_{-1.1}^{+1.4}$\,\unit{\kilo\parsec}}
\newcommand{\change}[1]{#1}
\newcommand{\changetwo}[1]{#1}

\usepackage{amsmath}
\usepackage{siunitx}
\usepackage{microtype}

\DeclareSIUnit{\mas}{mas}
\DeclareSIUnit{\mJy}{mJy}
\DeclareSIUnit{\beam}{beam}
\DeclareSIUnit{\wavelength}{\lambda}
\DeclareSIUnit{\utc}{UTC}
\DeclareSIUnit{\parsec}{pc}
\DeclareSIUnit{\GHz}{\giga\Hz}
\DeclareSIUnit{\year}{yr}

\begin{document}

\lefttitle{A Real-Time Jet Laboratory in \sjfull}
\righttitle{Wood et al.}

\jnlPage{x}{x}
\jnlDoiYr{2026}
\doival{10.1017/pasa.xxxx.xx}

\articletitt{Research Paper}

\title{A Real-Time Jet Laboratory in \sjfull}

\corresp{Callan M. Wood, Email: callan.wood@icrar.org}

\author{Callan M. Wood$^{1}$, James C. A. Miller-Jones$^{1}$, Arash Bahramian$^{1}$, Steven J. Tingay$^{1}$, Sara E. Motta$^{2}$, Hongmin Cao$^{3}$, Thomas D. Russell$^{4}$, Francesco Carotenuto$^{5,6}$, Pikky Atri$^{7,8}$, Diego Altamirano$^{9}$, Alexandra J. Tetarenko$^{10}$, Rob Fender$^{6}$, Elmar K\"{o}rding$^{8}$, Dipankar Maitra$^{11}$, Sera Markoff$^{12,13,14}$, David M. Russell$^{15}$, Gregory R. Sivakoff$^{16}$, Roberto Soria$^{17,18,19}$, and Valeriu Tudose$^{20}$}

\affil{$^1$International Centre for Radio Astronomy Research, Curtin University, GPO Box U1987, Perth, WA 6845, Australia}




\affil{$^2$INAF, Osservatorio Astronomico di Brera, Via E. Bianchi 46, I-23807 Merate, Italy}

\affil{$^3$School of Electronic and Electrical Engineering, Shangqiu Normal University, 298 Wenhua Road, Shangqiu, Henan 476000, People’s Republic of China}

\affil{$^4$INAF, Istituto di Astrofisica Spaziale e Fisica Cosmica, Via U. La Malfa 153, I-90146 Palermo, Italy}

\affil{$^5$INAF, Osservatorio Astronomico di Roma, Via Frascati 33, I-00078, Monte Porzio Catone (RM), Italy}
\affil{$^6$Astrophysics, Department of Physics, University of Oxford, Keble Road, Oxford, OX1 3RH, UK}

\affil{$^7$ASTRON, Netherlands Institute for Radio Astronomy, Oude Hoogeveensedĳk 4, 7991 PD Dwingeloo, The Netherlands}
\affil{$^8$Department of Astrophysics/IMAPP, Radboud University, P.O. Box 9010, 6500 GL, Nijmegen, The Netherlands}

\affil{$^9$School of Physics and Astronomy, University of Southampton, University Road, Southampton SO17 1BJ, UK}

\affil{$^{10}$Department of Physics and Astronomy, University of Lethbridge, Lethbridge, Alberta, T1K 3M4, Canada}



\affil{$^{11}$Department of Physics and Astronomy, Wheaton College, Norton, MA 02766, USA}

\affil{$^{12}$Anton Pannekoek Institute for Astronomy, University of Amsterdam, Science Park 904, 1098 XH Amsterdam, The Netherlands}
\affil{$^{13}$Gravitation and Astroparticle Physics Amsterdam Institute, University of Amsterdam, Science Park 904, 1098 XH 195 196 Amsterdam, The Netherlands}
\affil{$^{14}$Institute of Astronomy, University of Cambridge, Madingley Rd, Cambridge CB3 0HA, UK}

\affil{$^{15}$Center for Astrophysics and Space Science (CASS), New York University Abu Dhabi, P.O. Box 129188, Abu Dhabi, UAE}

\affil{$^{16}$Department of Physics, University of Alberta, CCIS 4-181, Edmonton AB T6G 2E1, Canada}

\affil{$^{17}$INAF, Osservatorio Astrofisico di Torino, Strada Osservatorio 20, 10025 Pino Torinese, Italy}
\affil{$^{18}$College of Astronomy and Space Sciences, University of the Chinese Academy of Sciences, Beijing 100049, People's Republic of China}
\affil{$^{19}$Sydney Institute for Astronomy, School of Physics A28, The University of Sydney, Sydney, NSW 2006, Australia}

\affil{$^{20}$Institute of Space Science - INFLPR Subsidiary, 077125 Magurele, Romania}

\citeauth{}

\history{(Received xx xx xxxx; revised xx xx xxxx; accepted xx xx xxxx)}

\begin{abstract}
Multi-wavelength observations of low-mass X-ray binaries (LMXBs) during bright outbursts reveal many details about the coupling of their inflows and outflows. However, only high angular resolution radio observations are able to resolve and track the motion and variability of individual jet ejecta. We present the results of our intensive VLBI campaign on the black-hole low-mass X-ray binary (LMXB) \sjfull\ during its 2023-2024 outburst. We observed the repeated quenching and re-establishment of the highly-extended continuous core jet during several transitions between hard-intermediate and soft-intermediate states, and the repeated ejection of transient jets. Using time-dependent visibility model fitting, we tracked the motion of nine discrete jet knots, obtaining some of the most precise measurements of transient jet proper motions and ejection dates in an LMXB. These ejecta were only detectable for a short time with VLBI, and some showed rapid intra-observation flux density variability that was not captured in image reconstructions. For the first time, we use time-dependent visibility modelling to fit a piecewise model for the jet knot flux densities, allowing us to create complex, non-parametric light curves of their intra-observation variability. \change{We observed the launching of multiple ejecta across several state transitions, however, we could not identify a consistent signature of jet ejection in the available X-ray intensity or hardness data.} We constrained the intrinsic speeds and bulk Lorentz factors of the jet knots, finding that \sjfull\ launched both mildly relativistic ($\beta\Gamma<$1) and highly relativistic ($\beta\Gamma>$2) ejecta throughout its outburst. We used their proper motions to constrain a posterior distribution for the maximum inclination angle of the jet axis, which had 50th, 84th, and 99th percentiles of \qty{40}{\degree}, \qty{50}{\degree}, and \qty{66}{\degree}, respectively. These unique observations of the repeated ejection of transient jets by a single LMXB reveal that fixed parameters such as black-hole mass, black-hole spin, and spin-orbit misalignment do not uniquely determine the varying properties of transient jets, particularly their speeds and Lorentz factors.
\end{abstract}

\begin{keywords}
Stellar mass black holes (1611), Radio jets (1347), Relativistic jets (1390), Very long baseline interferometry (1769), Low-mass X-ray binary stars (939), Transient sources (1851)
\end{keywords}

\maketitle

\section{Introduction} \label{sec:intro}
    During bright outbursts, the properties of the inflows and outflows in low-mass X-ray binaries (LMXBs) change dramatically on humanly accessible timescales, allowing us to probe the complex causal relationship between changes in the inner accretion flow and the launching of relativistic jets. During their outbursts, LMXBs transition through a series of accretion states, defined primarily by their X-ray spectral and timing properties \citep[see][for a review of black hole X-ray binary accretion states and state-transitions]{2006ARA&A..44...49R, 2010LNP...794...53B, 2016ASSL..440...61B, 2022hxga.book....9K}. In the canonical picture, LMXBs begin their outbursts in the hard state, where the X-ray spectrum is dominated by a hard Comptonised power-law component which exhibits strong variability. As the outburst progresses, the system transitions through a series of intermediate states and eventually reaches the  soft state, characterised by suppressed variability and a spectrum dominated by a soft thermal disk component. During the intermediate states, the X-ray spectral and timing properties evolve rapidly. Eventually, at the end of their outbursts, they undergo a reverse transition from the soft state back to the hard state, always at a lower luminosity than the original hard-to-soft transition. 

    The changes in the X-ray spectral and timing properties during the state transitions are also accompanied by changes in the properties of relativistic jets seen at lower frequencies \citep[see][for a review of accretion/ejection coupling in LMXB outbursts]{2004MNRAS.355.1105F, 2009MNRAS.396.1370F}. During the hard state, the radio emission is dominated by a compact self-absorbed optically thick continuous jet, which has a flat or slightly inverted spectrum extending from radio through sub-mm/mm and infrared wavelengths \citep[$\alpha\geq0,\ S_\nu\propto\nu^\alpha$;][]{2000A&A...359..251C, 2001MNRAS.322...31F, 2002ApJ...573L..35C, 2015ApJ...805...30T}, with a spectral break in the infrared wavelength range where the emission becomes optically thin \citep{2013MNRAS.429..815R}. Continuous jets have only been resolved in a handful of systems \citep{2000ApJ...543..373D, 2001MNRAS.327.1273S, 2004evn..conf..111R, 2015MNRAS.450.1745R, 2021MNRAS.504.3862T, 2024ApJ...971L...9W}, and hence they are often referred to as compact jets. As LMXBs transition from the hard state to the soft state, via intermediate states, the continuous jet generally quenches before it is re-established following the soft-to-hard reverse transition, which can be tracked through the evolution of the spectral break in broad-band SEDs \citep[e.g.][]{2013MNRAS.436.2625V, 2013MNRAS.431L.107C, 2014MNRAS.439.1390R, 2020MNRAS.498.5772R, 2024ApJ...962..116E}. During the transition through intermediate states, as the continuous jet quenches, optically thin ($\alpha\sim-0.7$) discrete transient ejecta are often launched. These ejecta can be tracked as they travel away from their launch site from milli-arcsecond to arcsecond scale distances \citep{2002Sci...298..196C, 2019ApJ...883..198R, 2020NatAs...4..697B, 2020ApJ...895L..31E, 2022MNRAS.511.4826C, 2023ApJ...948L...7B}, sometimes exhibiting apparently-superluminal motion \citep{1994Natur.371...46M, 1995Natur.375..464H, 1995Natur.374..141T, 2000ApJ...543..373D}. While this picture describes the broad evolution of LMXB outbursts, individual LMXBs can show a wide range of unique behaviours in their evolution through various accretion states \citep[see e.g.][and references therein]{2016ApJS..222...15T}, as well as in the properties and evolution of their relativistic jets. 
    
    The ejection of transient jets is often accompanied by radio flares and sudden changes in the X-ray spectral and timing properties of the inner accretion flow \citep{2001ApJ...554...43C, 2002MNRAS.331..765B, 2004MNRAS.347L..52G, 2004ApJ...617.1272C, 2012MNRAS.421..468M, 2017MNRAS.469.3141T}. While observations of transient jet knots at large distances from the core can be modelled to derive their ejection times and launch speeds \citep{2022MNRAS.511.4826C, 2024MNRAS.533.4188C, 2025MNRAS.tmp.1047C}, high angular resolution observations of the jet knots with very long baseline interferometry (VLBI) soon after they are launched are essential for tracking their motions and precisely inferring their ejection dates \citep{2012MNRAS.421..468M}. These observations, however, can be difficult to analyse due to the rapid motion and variability of these jet ejecta during an individual observations, violating a fundamental assumption of aperture synthesis. While imaging approaches such as time-binned (or snapshot) imaging \citep[e.g.][]{2001ApJ...558..283F, 2019Natur.569..374M} or proper motion correction techniques \citep[][]{2010MNRAS.409L..64Y, 2021MNRAS.505.3393W} are able to overcome some of these challenges, they can still struggle to create accurate reconstructions of these observations, particularly when the $uv$-coverage is sparse, when there are multiple fast-moving jet knots, and when those jet knots show rapid flux density variability. In \citet{2023MNRAS.522...70W}, we introduced a new analysis techniques called time-dependent visibility model fitting where we fit time-evolving models of the individual variable jet knots directly to the measured interferometric visibilities. This allows for their intra-observation motion, expansion, and flux density evolution to be directly measured, providing precise constraints on the transient jet properties and ejection dates from only a single high angular resolution observation \citep[e.g.][]{2025ApJ...984L..53W}. 
    
    Despite a gradually growing sample of LMXB outbursts with high angular resolution observations of transient ejecta together with dense, contemporaneous X-ray coverage, the precise connection between these changes in the X-ray timing properties and the launching and properties of transient ejecta is still unclear. One potential ejection signature is the switch from type-C to type-B quasi-periodic oscillations \citep[QPOs; see][for a review]{2019NewAR..8501524I} that occurs during the transition from the hard-intermediate state to the soft-intermediate state, which has now been observed close to the ejection times of transient jets in multiple LMXBs \citep{2020ApJ...891L..29H, 2021MNRAS.505.3393W}. There have been suggestions, however, that the appearance of type-B QPOs may not necessarily be a signature of ejection, since in some of these cases they have only been detected days apart from the inferred ejection dates, or not at all \citep{2012MNRAS.421..468M, 2019ApJ...883..198R, 2024MNRAS.533.4188C, 2026A&A...707A.151C}. The appearance of the type-B QPO has also been seen to occur in one source during a transition from a soft state back to a soft-intermediate state, simultaneously with infrared emission appearing from the compact jet \citep{2020MNRAS.495..182R}. Transient ejecta have never been observed to be launched during this reverse transition. There is no general consensus on the physical explanations of particular proposed ejection signatures.
    
   This is further complicated by the recent suggestion of a relationship between large-scale jet precession and the launch speeds of transient jet knots \citep{2025NatAs.tmp..198F}, which would imply that differing properties of transient jets, and in particular their intrinsic speeds, are closely related to the configuration of the inner accretion flow and the jet launching region. If the varying properties of transient jets are strongly linked to the properties and geometry of the inner accretion flow, then a single consistent signature of jet ejection might not be observed across the entire LMXB population, or even for multiple transient jets launched by the same LMXB, in the same outburst. 
   
   Therefore, high angular resolution observations of LMXBs in outburst, particularly of those that repeatedly launch transient jets, coupled with dense, contemporaneous X-ray coverage, are essential probes of jet launching. These observations are crucial for understanding the precise sequence of events in the inner accretion flow leading up to the ejection of transient jets, and the connection between the varied properties of these jets and the properties and geometry of the inner accretion flow and jet launching region. The extremely bright and well monitored 2023/2024 outburst of the black-hole LMXB \sjfull\ \citep[see][]{2025ApJ...988..109H} provided the perfect opportunity to study both the evolution of the continuous jet, and the repeated ejection of transient jets across an entire outburst with a dense high angular resolution VLBI campaign. 

    \subsection{\sjfull}\label{sec:Swift j1727 introduction}
        \sjfull\ was discovered on 2023 August 24 (MJD 60180) by \textit{Swift}/BAT \citep{2023GCN.34537....1P}. It was originally erroneously classified as a gamma-ray burst, however follow-up X-ray observations quickly identified it as a new candidate black hole low-mass X-ray binary in the hard-state at the beginning of a bright outburst \citep{2023GCN.34540....1K, 2023GCN.34544....1N, 2023ATel16205....1N, 2023ATel16206....1N, 2023GCN.34549....1O, 2023ATel16207....1O, 2023ATel16208....1C}. Following its detection, and in part due to its exceptional X-ray brightness (exceeding $7$~Crab in the 2--20\,keV band), an extensive multi-wavelength was launched to monitor its outburst. 

        Optical monitoring subsequently confirmed \sjfull\ to be LMXB containing a black hole in a $\sim$\qty{10}{\hour} binary orbit with (most likely) an early K-type dwarf companion \citep{2025A&A...693A.129M}. Based on HI absorption spectra and near-UV colour excess, \citet{2025ApJ...994..243B} inferred a distance to \sjfull\ of \distanceBurridge, assuming that the companion is an unevolved K4($\pm1$)V main sequence star, noting that significant evolution of the companion would imply a closer distance. 
        
        X-ray monitoring revealed that as \sjfull\ progressed further into the hard-intermediate state, it entered a period of rapid X-ray flaring \citep[a so called `flaring state'; see e.g.][]{2024arXiv240603834L, 2024ApJ...970L..33Y, 2024MNRAS.529.4624Y, 2024ApJ...974..303Z, 2025ApJ...986....3L}, beginning around 2023 September 10 (MJD 60197). This period was characterised by rapid fluctuations in the X-ray intensity, spectral, and timing properties due to rapid changes in the emission and geometry of the inner accretion disk and the Comptonised corona \citep{2025ApJ...986....3L, 2025ApJ...983...23C, 2025arXiv250801384H, 2026A&A...706A.208J}. 
        
        \sjfull\ was reported to have briefly transitioned between the hard-intermediate and soft-intermediate states on multiple occasions, including a very short-lived transition during a bright X-ray flare during the flaring state \citep{2026A&A...706A.208J}. This flaring period eventually ended when \sjfull\ underwent a prominent transition from the hard-intermediate state to the soft-intermediate on 2023 October 5 \citep[MJD 60222;][]{2023ATel16273....1B, 2023ATel16271....1M}, followed by a reverse transition and then a second prominent hard-intermediate to soft-intermediate state transition on 2023 October 14 \citep[MJD 60231;][]{2023ATel16276....1Y, 2023ATel16289....1T}, before it finally progressed towards the soft state. Due to the prominence of the hard-intermediate to soft-intermediate state transitions on 2023 October 5 and 2023 October 14 \citep[i.e. that they were identified and reported shortly after they occurred;][]{2023ATel16273....1B, 2023ATel16271....1M, 2023ATel16276....1Y, 2023ATel16289....1T}, we refer to these throughout this paper as the first and second prominent hard-intermediate to soft-intermediate state transitions, respectively.

        We note that the `flaring state' is not a canonical accretion state. Based on spectral modelling, \citet{2025ApJ...983...23C} suggested that from MJD 60197 to MJD 60204, \sjfull\ was in the process of transitioning from a hard-intermediate state to a very high state, where it remained for the duration of their observations (until MJD 60220). However, based on timing studies during the period of MJD 60204 to MJD 60209, which showed the presence of a strong type-C QPO, \citet{2026A&A...706A.208J} identified \sjfull\ as being in the hard-intermediate state. Throughout the rest of this paper, we refer to the period between MJD 60197 and MJD 60222 as the flaring state. A detailed analysis of the X-ray spectral and timing observations and the precise state classifications during this period is beyond the scope of this paper.
        
        Along with dense X-ray observations, \sjfull\ was comprehensively monitored at radio wavelengths throughout the outburst, showing multiple periods of radio quenching and optically thin flaring coincident with changes in the X-ray light curves, including two long-lived radio flares following the two most prominent hard-intermediate to soft-intermediate state transitions \citep[][see also Figure~\ref{fig:MAXI Lightcurve Observation Summary}]{2025ApJ...988..109H}. While these radio flares likely signified the ejection of transient jets that eventually propagated out to arcsecond scale distances (A. Hughes, priv. comm., 2025), only high angular resolution observations can be used to track their motions close to the core and infer their precise ejection dates to constrain the complex sequence of events leading up to the launching of transient ejecta. 

        In this paper we present the results of our comprehensive high angular resolution VLBI observing campaign of \sjfull\ during its 2023-2024 outburst. In Section~\ref{sec:campaign} we describe our campaign, and in Section~\ref{sec:VLBI analysis} we detail our calibration, imaging, and visibility modelling analysis procedures. In Section~\ref{sec:results} we present the imaging and visibility modelling results for each observation, and we summarise the parameters of the detected and modelled transient jet knots. Finally, in Section~\ref{sec:discussion}, we discuss the properties of the continuous and transient jets launched by \sjfull, as well as their evolution throughout the outburst and their connection to X-ray and other radio observations.

\section{VLBI Observing Campaign}\label{sec:campaign}
\begin{figure*}
        \centering
        \includegraphics[width=\linewidth]{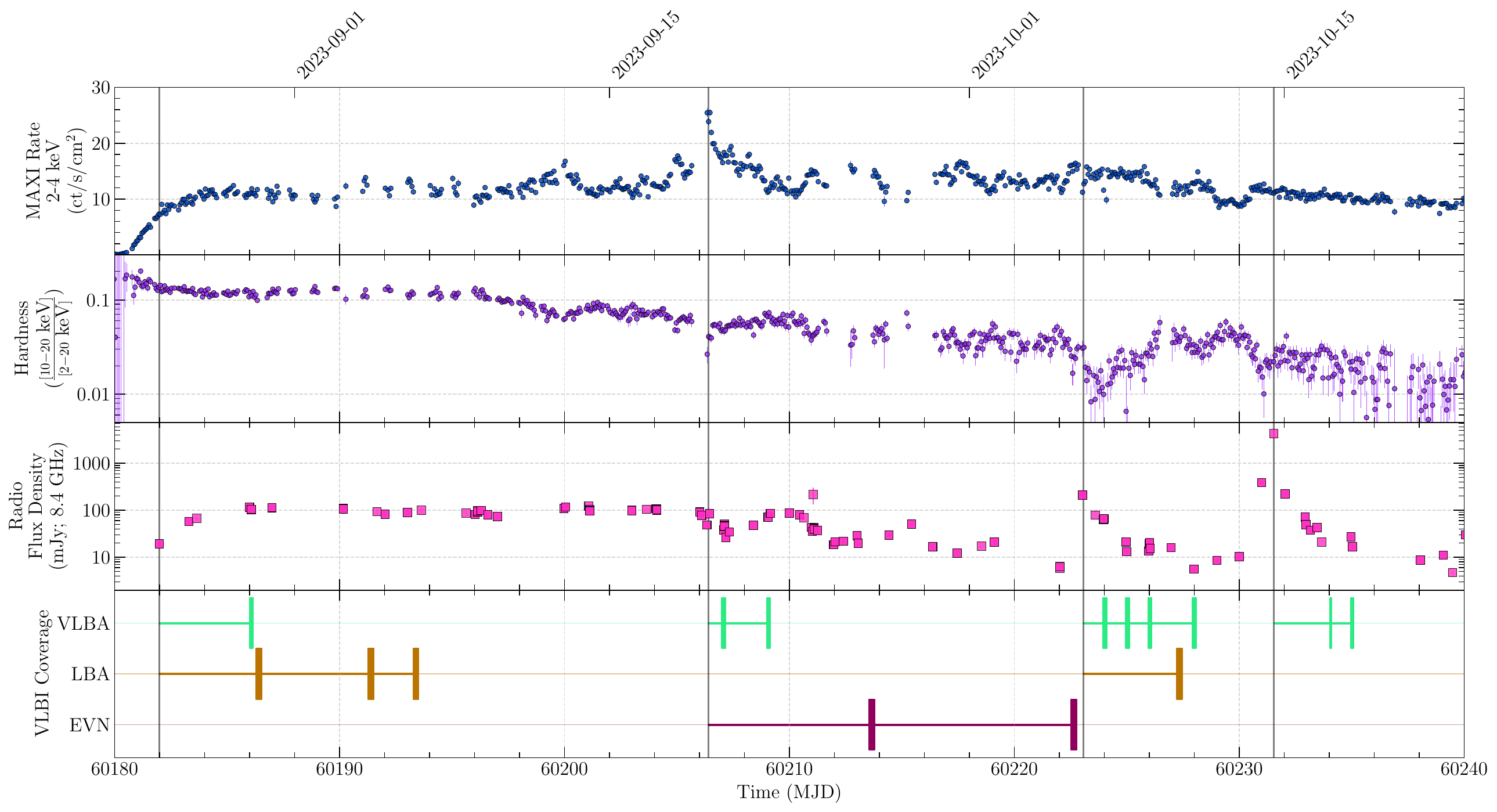}
        \caption{Overall X-ray and radio evolution of \sjfull\ and the 15 VLBI observations taken during the peak of the 2023 outburst. The top panel shows the 2-4 keV MAXI/GSC\textsuperscript{a} X-ray light curve \citep{2009PASJ...61..999M}, the second panel shows the MAXI/GSC X-ray hardness ratio, and the third panel shows the overall radio light curves presented in \citet{2025ApJ...988..109H}, re-scaled to \qty{8.4}{\GHz}. The vertical coloured bars in the bottom panel show the times and durations of the VLBI observations, and they are connected via the horizontal coloured lines to the vertical black lines, which show the timing of the respective events that triggered our observations. These triggers were: the first radio detection of \sjfull\ on 2023 August 26 \citep[MJD 60182;][]{2023ATel16211....1M}, the bright X-ray flare on 2023 September 19 (MJD 60206), the radio flare detected on 2023 October 6 \citep[MJD 60223;][]{2023ATel16271....1M}, and the radio flare detected on 2023 October 14 \citep[MJD 60231;][]{2023ATel16289....1T}. The first four VLBI observations were published in \citet{2024ApJ...971L...9W}, the fifth was published in \citet{2025ApJ...984L..53W}, and the second EVN observation was published in \citet{2025ApJ...987L..14C}. We took a final astrometric observation with the VLBA on 2024 March 24 (MJD 60393) following the reverse transition to the hard-state \citep{2024ATel16541....1P, 2024ATel16552....1R}, which is not included \change{here}.}
        \raggedright{\footnotesize\textsuperscript{a}\url{http://maxi.riken.jp/}}
        \label{fig:MAXI Lightcurve Observation Summary}
    \end{figure*}

    We summarise our full VLBI observing campaign of \sjfull\ in Figure~\ref{fig:MAXI Lightcurve Observation Summary} and Table~\ref{tab:vlbi observation log}, which consisted of 10 observations with the Very Long Baseline Array (VLBA), four with the Long Baseline Array (LBA), and two with the European VLBI Network (EVN). We have divided the observations into the various phases of the outburst, which we detail below, including our previously published observations \citep[see][]{2024ApJ...971L...9W, 2025ApJ...984L..53W, 2025ApJ...987L..14C}.

    \begin{table*}
        \begin{center}
        \caption{\sjfull\ full VLBI Observing Campaign Log.}
        \label{tab:vlbi observation log}
        {\tablefont
        \begin{tabular}{@{\extracolsep{\fill}}llccclccc}
            \toprule
           Code & Date & State\textsuperscript{a} & Time & Midpoint & Telescope & Frequency & Bandwidth & Stations\textsuperscript{b,c}\\
                           &                &                 & (UTC)  & (MJD)              &                     & (GHz)   & (MHz)   & \\
            \hline
            BM538A\textsuperscript{d,e} & 2023~Aug~30    & HIMS & 00:33--03:21   & 60186.08 & VLBA & 8.37          & 512     & FD,HN,KP,LA,MK,NL,OV,SC          \\
            V456H\textsuperscript{e}      & 2023~Aug~30    & HIMS & 07:06--12:47   & 60186.41 & LBA  & 8.44          & 64      & CD,HO,KE,MP,PA,WW                \\
            V456I\textsuperscript{e}      & 2023~Sep~04    & HIMS & 06:36--12:20   & 60191.39 & LBA  & 8.44          & 64      & AT,CD,HO,KE,MP,PA,WW             \\
            V456J\textsuperscript{e}      & 2023~Sep~06    & HIMS & 06:36--12:20   & 60193.39 & LBA  & 8.44          & 64      & AT,CD,MP,PA,WW                   \\
            BM538B\textsuperscript{f}     & 2023~Sep~19/20 & FS   & 23:56--03:41   & 60207.08 & VLBA & 2.33/8.37\textsuperscript{g} & 128/384 & FD,KP,LA,MK,NL,OV,PT,SC          \\
            BM538C\textsuperscript{h}     & 2023~Sep~22    & FS   & 00:07--03:30   & 60209.08 & VLBA & 8.37          & 512     & BR,FD,KP,LA,MK,NL,OV,PT,SC       \\
            RM018          & 2023~Sep~26    & FS   & 13:21--18:51   & 60213.67 & EVN  & 4.93          & 256     & JB,WB,EF,MC,O8,TR,IB,DA,KN,PI    \\
            RM019\textsuperscript{i}      & 2023~Oct~05    & SIMS & 12:49--18:23   & 60222.65 & EVN  & 4.93          & 256     & JB,WB,EF,MC,O8,HH                \\
            BM538D         & 2023~Oct~06/07 & SIMS & 23:08--02:31   & 60224.03 & VLBA & 8.37          & 512     & BR,FD,HN,KP,MK,NL,PT             \\
            BM538E         & 2023~Oct~07/08 & SIMS & 23:04--02:12   & 60225.03 & VLBA & 8.37          & 512     & BR,FD,HN,KP,MK,NL,PT             \\
            BM538F         & 2023~Oct~08/09 & SIMS & 23:00--02:23   & 60226.03 & VLBA & 8.37          & 512     & BR,FD,HN,KP,LA,MK,NL,PT          \\
            V456K          & 2023~Oct~10    & HIMS & 05:06--10:50   & 60227.33 & LBA  & 8.44          & 64      & AT,HO,MP,WW                      \\
            BM538G         & 2023~Oct~10/11 & HIMS & 22:04--01:49   & 60228.00 & VLBA & 2.33/8.37\textsuperscript{g} & 128/384 & BR,FD,HN,KP,LA,MK,NL,PT          \\
            BM538H         & 2023~Oct~17    & SIMS & 00:39--02:24   & 60234.06 & VLBA & 8.37          & 512     & BR,FD,HN,KP,LA,MK,NL,OV,PT       \\
            BM538I         & 2023~Oct~17/18 & SIMS & 23:17--01:47   & 60235.02 & VLBA & 8.37          & 512     & BR,FD,HN,KP,LA,MK,NL,OV,PT       \\
            BM538J\textsuperscript{d}     & 2024~Mar~24    & HS   & 11:10--13:58   & 60393.52 & VLBA & 8.37          & 512     & BR,FD,KP,LA,MK,NL,OV,PT          
            \botrule
        \end{tabular}}
        \end{center}
        \begin{tabnote}
            {\textsuperscript{a} HIMS: hard-intermediate state; FS: flaring state; SIMS: soft-intermediate state; and HS: hard state}\tnp
            {\textsuperscript{b} We only list stations that successfully recorded data and were not flagged out due to poor conditions and/or failed calibration.}\tnp
            {\textsuperscript{c}VLBA Stations: BR=Brewster, FD=Fort Davis, HN=Hancock, KP=Kitt Peak, LA=Los Alamos, MK=Mauna Kea, NL=North Liberty, OV=Owens Valley, PT=Pie Town, SC=St. Croix; EVN Stations: JB=Jodrell Bank (Mk II), WB=Westerbork, EF=Effelsberg, MC=Medicina, O8=Onsala, TR=Torun, HH=Hartebeesthoek, IB=Irbene, DA=Darnhall, KN=Knockin, PI=Pickmere; LBA Stations: CD=Ceduna, HO=Hobart 12m, KE=Katherine, MP=Mopra, PA=Parkes, WW=Warkworth 30m, AT=ATCA.}\tnp
            {\textsuperscript{d} Included geodetic blocks \citep{2009ApJ...700..137R} for improved astrometric calibration.}\tnp
            {\textsuperscript{e} Published in \citet{2024ApJ...971L...9W}.}\tnp
            {\textsuperscript{f} Published in \citet{2025ApJ...984L..53W}.}\tnp
            {\textsuperscript{g} Observed with the VLBA S/X-band dichroic feed.}\tnp
            {\textsuperscript{h} Configured for polarisation calibration (not included in this publication).}\tnp
            {\textsuperscript{i} Published in \citet{2025ApJ...987L..14C}.}\tnp
        \end{tabnote}
    \end{table*}

    \subsection{Hard/Hard-Intermediate State}
        Following the initial X-ray discovery \citep{2023GCN.34537....1P, 2023GCN.34540....1K} and subsequent radio detection of \sjfull\ \citep{2023ATel16211....1M}, we triggered our VLBA observing campaign (project code BM538) as part of the Jet Acceleration and Collimation Probe Of Transient X-Ray Binaries \citep[JACPOT XRB;][]{2011IAUS..275..224M} program. At the same time, we also triggered our LBA observing program (project code V456). We initially took one VLBA and three LBA observations of the source in the hard/hard-intermediate state between 2023 August 30 and 2023 September 6 (MJD 60186-60193), which we published in \citet{2024ApJ...971L...9W}. These observations revealed a bright core, and a large, two-sided, asymmetrical resolved jet extending in the north-south direction. The images showed the most well-resolved continuous X-ray binary jet to date, and they showed that \sjfull\, may have had, depending on its distance, the most physically extended continuous jet ever seen in an X-ray binary. In the first VLBA observation, we also detected an apparently disconnected transient jet knot travelling to the south \citep[Figure~2 of][]{2024ApJ...971L...9W}, which for the remainder of this paper we will refer to as `knot 0'. While we were able to track the proper motion of this apparently discrete jet knot, in \citet{2024ApJ...971L...9W} we interpreted it as being the result of a jet-ISM interaction or a downstream internal shock, rather than as a transient relativistic jet launched during the rising hard/hard-intermediate state.

    \subsection{Flaring State}
        Two weeks after our third LBA observation, we scheduled another observation with the VLBA to study the multi-frequency properties of the extended continuous jet, and to look for evidence of transient jet launching. For these reasons, we scheduled the observation using the dichroic S/X-band feed that simultaneously observes at 2.3 and \qty{8.3}{\GHz}. This observation was taken during the flaring state, and it occurred immediately following the bright, soft, X-ray flare that began on 2023 September 19 (MJD 60206), which was the brightest X-ray flare of the entire outburst (see Figure~\ref{fig:MAXI Lightcurve Observation Summary}). In \citet{2025ApJ...984L..53W} we described how we were able to track the motion, flux density evolution, and expansion of three discrete jet knots, which we labelled `knot 1', `knot 2', and `knot 3' \citep[see Figure~2 of][]{2025ApJ...984L..53W}. We also detected the compact core, and by measuring the frequency-dependent shift in its position between 2.3 and 8.3~GHz, we were able to determine the approximate location of the central black hole. Using these measurements, and the relative motions of the three distinct jet knots, we were able to show that they were all likely launched during the bright X-ray flare, which coincided with a change in the X-ray timing properties of the inner accretion flow, and was potentially a short-lived transition between the hard-intermediate and soft-intermediate states \citep{2026A&A...706A.208J}. 
          
        We took one more VLBA observation of \sjfull\ during the flaring state on 2023 September 22 (MJD 60209). Given the brightness of \sjfull, we configured the third VLBA observation with full polarisation calibration, however we only explore the Stokes I results in this paper. Following this observation, we obtained two target of opportunity observations with the EVN to continue to study the evolution of the jets of \sjfull, the first of which was taken during the flaring state on 2023 September 26 (MJD 60213). The EVN observations were taken at a central frequency of \qty{4.93}{\GHz}, unlike the LBA and VLBA observations, which were centred in the \qty{8}{\GHz} band. 

    \subsection{First State-Transition}
        \citet{2023ATel16271....1M} reported that at 00:27 UTC on 2023 October 5 (MJD 60222), the compact jet flux density of \sjfull\ had quenched to a level of \qty{7.2\pm0.1}{\mJy} at \qty{5.25}{\GHz}, but was then detected at \qty{235.8\pm1.1}{\mJy} the following day at 01:25 UTC, which suggested the launching of transient ejecta. \citet{2023ATel16273....1B} subsequently reported that NICER observations indicated that \sjfull\ had undergone the hard-intermediate to soft-intermediate state transition sometime between 03:00-11:00 UTC on 2023 October 5 (MJD 60222), prior to the reported radio flare. Our second EVN observation of \sjfull\ was serendipitously taken within hours of the NICER-reported state-transition, prior to the VLA-detected radio flare. This EVN observation was analysed by \citet{2025ApJ...987L..14C}, who showed that it contained rapid amplitude variability on both the long and short baselines (i.e. on both small and large size-scales). This amplitude variability, combined with the sparse $uv$-coverage, made imaging almost impossible, and thus they could only argue that the intra-observation variability of \sjfull\, was due to the ejection of at least one transient jet knot. This scenario was supported by simulations of the observation containing a fixed core and a moving transient jet knot, although they were unable to fully capture the complex behaviour of the visibility amplitudes, and they did not discuss the time-varying visibility phases.

        Following the reported state-transition and radio flaring, we observed four times with the VLBA at an almost daily cadence from 2023 October 6 to 11 (MJD 60223-60228), and once with the LBA, to track the motions of any jet knots launched during the state transition. This was our final LBA observation in the campaign. The daily VLBA observations were plagued by poor weather conditions and technical difficulties at a number of stations, which made calibration challenging. 
        
    \subsection{Second State-Transition}
        \cite{2023ATel16276....1Y} reported that in a MAXI observation on 2023 October 11 (MJD 60228), \sjfull\ had a similar hardness ratio as before the 2023 October 5th state transition, suggesting that it had returned to the hard-intermediate state (see Figure~\ref{fig:MAXI Lightcurve Observation Summary}). On 2023 October 14 (MJD 60231), the RATAN-600 telescope detected a second bright radio flare from \sjfull, suggesting the ejection of more transient relativistic ejecta following another hard-intermediate to soft-intermediate state transition \citep{2023ATel16289....1T}, prompting us to schedule two more observations with the VLBA on 2023 October 17/18 (MJD 60234/60235). Due to its detection at the beginning of the weekend, there was a few-day delay between the peak of the flare and our observations. Following this, \sjfull\ eventually reached the soft state, where it remained until the eventual soft-to-hard reverse transition in March 2024.

    \subsection{Reverse Transition}
        Finally, after the eventual reported soft-to-hard reverse transition and radio detection in March 2024 \citep{2024ATel16541....1P, 2024ATel16552....1R}, we observed once more with the VLBA to track the proper motion (and potentially the parallax) of the compact core. As with the first VLBA observation \citep{2024ApJ...971L...9W}, we observed with geodetic blocks \citep{2009ApJ...700..137R} for $\sim30$ minutes at the beginning and end of the observation to improve the tropospheric calibration for more precise astrometry \citep{2009ApJ...700..137R}. This was our final observation of \sjfull, and it has since returned to quiescence where it is too faint to detect.
    
\section{VLBI Data Analysis}\label{sec:VLBI analysis}
    \subsection{Calibration and Imaging}
        For the VLBA observations, we observed a combination of J1743-0350, J1642+3948, J1924-2914, J2253+1608, and J2136+0041 as fringe finders. We used the nearby (1.50\unit{\degree} separation) ICRF J172134.6-162855 (J1721-1628) as a phase reference source, and the slightly further (1.63\unit{\degree} separation) ICRF J172446.9-144359 (J1724-2914) as a check source \citep{2020A&A...644A.159C}. For the LBA observations, we observed a combination of J1733-1304, 1921-293, and 3C273 as fringe finders, and we swapped the phase reference and check sources for the LBA to maximise the signal-to-noise on the long baselines for phase calibration. For the EVN, we observed J1743-0350, J2136+0041, 3C286, and OQ208 as fringe finders, and used the same phase reference source and check source as the LBA. \change{All of the VLBA and LBA data were correlated using the DiFX software correlator \citep{2007PASP..119..318D, 2011PASP..123..275D}, and calibrated according to the standard procedures within the Astronomical Image Processing System \citep[\aips, version 31DEC22;][]{1985daa..conf..195W, 2003ASSL..285..109G}. The EVN data were correlated using the SFXC correlator at the Joint Institute for VLBI ERIC (JIVE). The amplitude and bandpass calibration was performed using the EVN calibration pipeline\footnote{\url{https://www.evlbi.org/evn-pipeline}} following the EVN Data Reduction Guide\footnote{\url{https://www.evlbi.org/evn-data-reduction-guide}}, but we re-performed the phase calibration within \aips.} For the joint 2.3 and \qty{8.3}{\GHz} VLBA observations (BM538B and BM538G), we split the the data into four 128-MHz intermediate frequency (IF) pairs, with the lowest IF pair centered at \qty{2.3}{\GHz} and the other three IF pairs centered at \qty{8.3}{\GHz}, and calibrated the bands individually. After the standard external gain calibration, we performed several rounds of imaging and self-calibration (hybrid mapping) of the phase reference source to produce the best model of the source, to then derive the most accurate time-varying phase, delay, and rate solutions, which we interpolated to \sjfull. We also performed a single round of amplitude self-calibration to get the most accurate time-varying amplitude gain calibration, which we applied to \sjfull. To match the flux density scales of the VLBA and the LBA, we used the VLBA map of J1724-2914 to derive a global amplitude gain solution for each LBA antenna, IF, and polarisation to scale the a priori amplitude gains approximated from the zenith system equivalent flux densities. We used the VLBA and LBA scans of J1721-1628 to confirm that the LBA flux density scale matched the VLBA to within 5\%. We adopt a 10\% calibration uncertainty for the VLBA, and 20\% uncertainty for the LBA and EVN, which we add in quadrature with the statistical uncertainty in any reported flux densities. 

        While many of our observations contained rapidly variable jets that we eventually studied with time-dependent visibility modelling, we first had to perform traditional static imaging to explore the data. \change{As discussed in Section~\ref{sec:intro}, traditional imaging techniques like CLEAN \citep{1974A&AS...15..417H, 1978A&A....65..345S, 1980A&A....89..377C} or static visibility model fitting \citep[e.g.][]{1994BAAS...26..987S} are designed to reconstruct a static model of the underlying source structure from the visibilities of an entire observation. They are not well suited to reconstruct observations of time-varying sources like transient jets \citep[see][and references therein]{2024evn..conf...25M}. Despite their limitations, we still use these methods to form an initial impression of the general source morphology to motivate our time-variable model selection and choice of priors.} We performed our initial imaging of \sjfull\ with both \aips\ and \difmap. We first imaged in \aips, which involved several rounds of CLEANing and self-calibration. For observations where a compact (or extended) core was detected, we used the \aips\ task \texttt{JMFIT} to fit the location of the peak of the core in the image plane. We also performed several rounds of imaging using Gaussian tapering, where we down-weighted the long baselines to increase our sensitivity to larger scale structure. This allowed us to detect multiple diffuse or smeared out jet knots \citep[see e.g.][]{2025ApJ...984L..53W}. For observations where we detected distinct transient jet knots, or where the core was particularly compact, we used \difmap\ to model the isolated knots with either circular or elliptical Gaussian components. The \difmap\ images contain a combination of static model-fit components for the compact knots, and clean components for the extended core, and they were created with several rounds of CLEANing, model-fitting, and self-calibration. In some observations, the amplitude variability of the source made it impossible to image the observation in full, and so we produced images from several small segments of the observations where the source was static.

    \subsection{Time-dependent Model Fitting}
        Many of the observations contained distinct transient jet knots. To characterise their intra-observation motions, expansions, and flux density evolutions, we use the time-dependent visibility model-fitting method that was first described in \citet{2023MNRAS.522...70W}. \change{In this method, we fit analytical model components directly to the externally calibrated visibilities, much like in \difmap, however we allow these models to vary in time throughout the observation, and we parameterise that variability explicitly \citep[see][for more details and for comparisons with traditional imaging techniques]{2023MNRAS.522...70W, 2024ApJ...971L...9W, 2025ApJ...984L..53W}}. We used the iterative self-calibration procedure implemented in \citet{2025ApJ...984L..53W}, where we perform several iterative rounds of phase-only self-calibration and model fitting, to derive our best time-varying model of the observation. In some observations, we found that we could not adequately represent the amplitude variability with a simple monotonic function. For the first time, we fit the flux density evolution of the individual components with a piecewise linear light curve, where we solved for their flux density at a series of evenly spaced fixed times (the number of which are specified in the model) and linearly interpolate the flux density between those times. In one case (observation BM538D), we fit for the flux density of a jet knot in each individual scan. This allows us to track the complex non-parametric evolution of the flux density of the jet knots.

        For our modelling, we used the Bayesian inference algorithm Nested Sampling \citep{10.1214/06-BA127}, implemented in the \textsc{dynesty}\footnote{\url{https://github.com/joshspeagle/dynesty}} Python package \citep{2020MNRAS.493.3132S}. We assume that the noise on the complex visibilities is circularly Gaussian, and so we use a Gaussian likelihood. \change{We used the results of our static \difmap\ model fitting to infer reasonable bounds on the fitted parameters of individual jet components, such as their sizes and their locations within an observation. We placed uniform priors on these parameters within those bounds. Restricting the bounds of our priors based on the results of our initial imaging analysis allowed for more efficient sampling of the posterior, which was useful when exploring different model configurations. However, we also re-performed many of our model fits with relaxed prior constraints to ensure that we were confident that the sources that we were fitting to \changetwo{corresponded to genuine astrophysical emission rather than thermal noise peaks, calibration errors, or imaging/modelling artefacts,} and that our results were insensitive to our choice of priors. As we show in Section~\ref{sec:results} and as we discuss in Section~\ref{sec:position angle and precession}, we found that the discrete jet knots launched by \sjfull\ were aligned along a fairly constant position angle. This is consistent with multi-wavelength studies that showed that the position angle of the jet axis and inner accretion disk in \sjfull\ was relatively stable and consistent throughout the entire outburst \citep[][]{2024ApJ...971L...9W, 2024ApJ...966L..35S, 2024ApJ...968...76I}. We therefore chose a Gaussian prior for the orientation of elliptical Gaussian model components and the direction of motion of transient jets knots with a mean of \qty{180}{\degree} East of North and a standard deviation of \qty{2}{\degree} \citep{2024ApJ...971L...9W}, unless otherwise stated.}

        For each observation, we tested a range of models, including with both circular and Elliptical Gaussian components, with and without linear expansion, and with varying number of piecewise linear segments in the light curves. We also tested models with multiple discrete jet components, however we found that only the second VLBA observation \citep[BM538B;][]{2025ApJ...984L..53W} contained multiple discrete jet knots. \change{For each observation, we selected our best fitting model based on a range of factors. Our first step was to discard models that did not produce physically reasonable results. Examples of models that we discarded were: models that produced components that were too large and diffuse to be detected by the telescope, models where jet components moved perpendicular to the known direction of the jet axis, or models where the expansion was so rapid that the component had zero size at some point during the observation.} We also inspected plots comparing the model visibilities to the measured visibilities, as well as images created from the model-subtracted visibilities (i.e. residual images) to look for any significant unmodelled structure. \change{This also allowed us to confirm that any additional components that we fit to the observation correspond with significant structure in the residual visibilities and residual image, and that we were not fitting components to peaks caused by thermal noise or calibration errors.}  This method is common in traditional imaging and model fitting procedures, where additional CLEAN or model components are included in regions with large residual flux density. Inspecting the residuals is also useful for motivating rounds of phase-only or phase and amplitude self-calibration, as errors in the calibration manifest in the residual maps. We also often found that apparent amplitude calibration errors could be resolved by including model components with a time-variable flux density, and apparent phase and amplitude errors surrounding a fixed component can sometimes hint towards intrinsic motion and amplitude variability that should be included in the model. \change{In our iterative model-fitting and self-calibration approach, we always began with the simplest model configurations, and only added extra complexity to the model if there was evidence of unmodelled structure in the residual visibilities and residual image.} \change{To confirm that our fitted results were reliable, we divided individual observations into shorter segments which we modelled individually and compared with the results from modelling the entire observations in full.} 

        \change{Finally, we used the Bayesian evidence provided by nested sampling to compute the ratio of posterior evidences (the Bayes factor) to compare models with similar but slightly different parameterisations. This was most useful for determining an appropriate number of segments in the piecewise linear light curve models to ensure that we were not over fitting the data. In some cases, however, the Bayes factor could not clearly discriminate between very similar models, for example: models with 5 or 6 piecewise segments; or between an elliptical Gaussian model where the major and minor axis were only slightly different in size compared to a circular Gaussian model. In these cases, we chose the model configuration with fewer free parameters. As we already noted, we included a 10\% systematic uncertainty in all flux density measurements. A consequence of this was that piecewise light curve models with different numbers of piecewise segments (e.g. 5 segments vs 6 segments) often yielded results that were consistent within these inflated uncertainties.}

        For each observation that we successfully modelled, we report the median of the marginal probability distributions as the best fit values of the model parameters, and the 16th and 84th percentiles as the uncertainties. For some model parameters, we include additional systematic uncertainty, which we describe in more detail further in the text. 
        
    \begin{figure*}
        \centering
        \includegraphics[width=\linewidth]{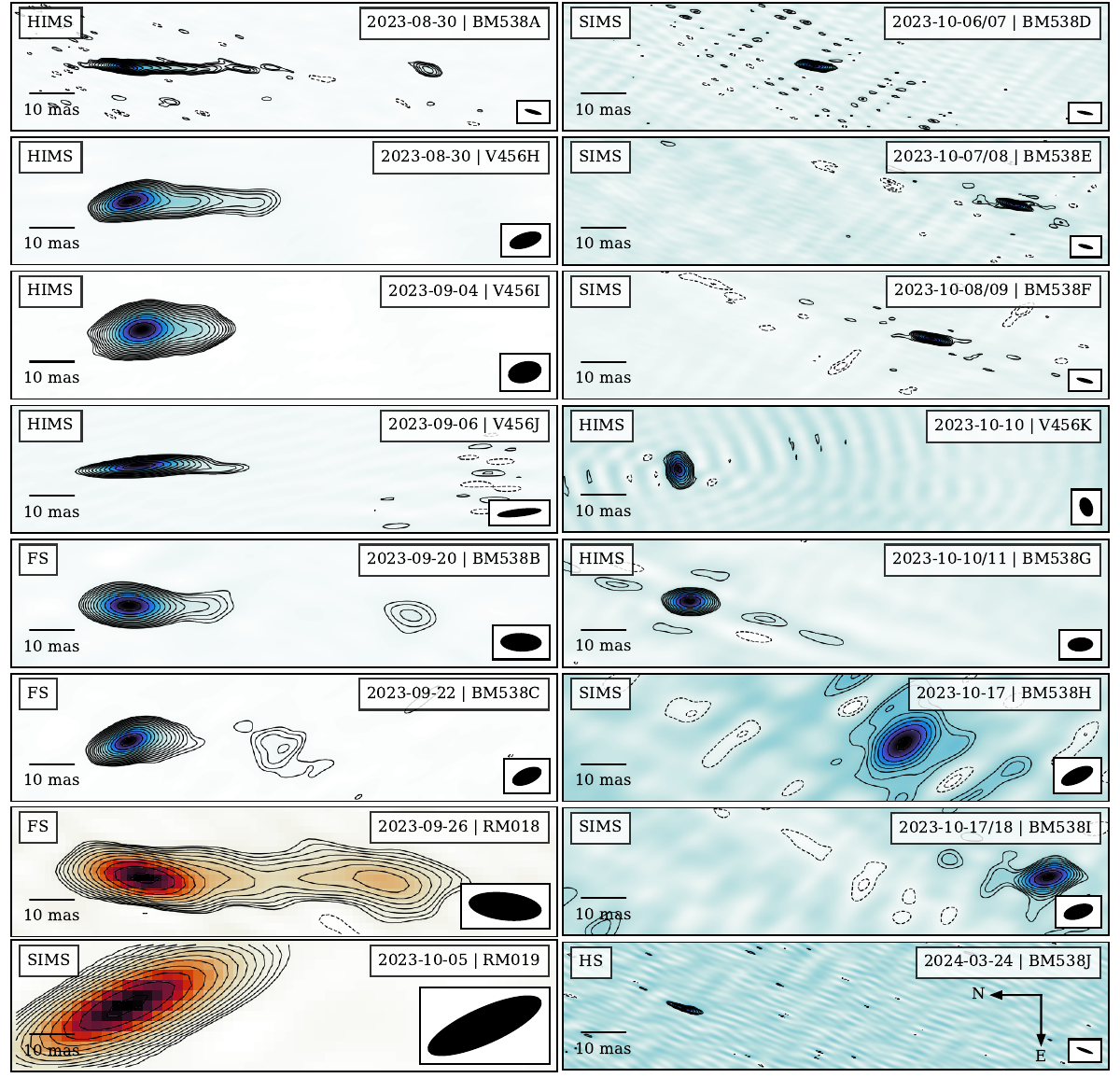}
        \caption{Montage of images from our VLBI campaign during the 2023/2024 outburst of \sjfull. Time progresses down the left column and then down the right column. These static reconstructions show the evolution of the continuous jet and the ejection of multiple transient jet knots. See Table~\ref{tab:vlbi observation log} for observation details. The images have been rotated \qty{90}{\degree} counter-clockwise, as marked by the compass in the lower right panel. The colour scale and contours are described in Figures~\ref{fig:bm538c}, \ref{fig:rm018}, \ref{fig:state transition montage}, \ref{fig:bm538I bm538H images}, \ref{fig:bm538j}, in Figure~3 of \citet{2024ApJ...971L...9W} and Figure~1 of \citet{2025ApJ...984L..53W}. The restoring beams are shown by the black ellipses. The labels in the top right give the dates and observation codes, and the labels in the top left show the accretion state of \sjfull\ during the observations\textsuperscript{a}. Observations at \qty{8.4}{\GHz} are in blue and observations at \qty{4.9}{\GHz} are in red. Observations BM358B and BM538G have simultaneous \qty{2.3}{\GHz} data, which we do not show in this montage (see Figure~1 of \citet{2025ApJ...984L..53W}, and Figure~\ref{fig:state transition montage} here). The images of RM019 and BM538D were made from only a short slice of the data (see Sections~\ref{sec:rm019} and \ref{sec:BM538D, BM538E, BM538F, V456K, and BM538G}). We show the evolution of the position of the core (corrected for small systematic astrometric offsets) in Figure~\ref{fig:core astrometry}.}
        \raggedright{\footnotesize\textsuperscript{a} HIMS: hard-intermediate state; FS: flaring state; SIMS: soft-intermediate state; and HS: hard state}
        \label{fig:Full image montage}
    \end{figure*}

\section{Results}\label{sec:results}
    In Figure~\ref{fig:Full image montage}, we show a montage of images made from each observation of \sjfull. All of these images (and all of the images we show in this paper) are static reconstructions of the observations, and thus they do not capture the motion and time-variability of the jets launched by \sjfull. \change{For each observation in Figure~\ref{fig:Full image montage}, we include the observation codes by which we refer to them for the remainder of this paper (see also Table~\ref{tab:vlbi observation log}).} The first five images in this montage have previously been published in \citet{2024ApJ...971L...9W} and \citet{2025ApJ...984L..53W}. We do not include the simultaneous \qty{2.3}{\GHz} images from observations BM538B and BM538G in this \change{Figure}. The montage shows that over the course of the outburst, we saw both the evolution of the continuous extended core jet, as well as the presence of multiple discrete jet knots. While these images show the overall evolution of the jets of \sjfull\ during its 2023-2024 outburst, they do not capture the intra-observation variability and motion of the jet knots. We now present static imaging and time-dependent visibility modelling results for the unpublished observations from our campaign, beginning with the remaining observations of the flaring state. 

    \subsection{Flaring State}
        \subsubsection{BM538C}
\begin{figure}
                \centering
                \includegraphics[width=\linewidth]{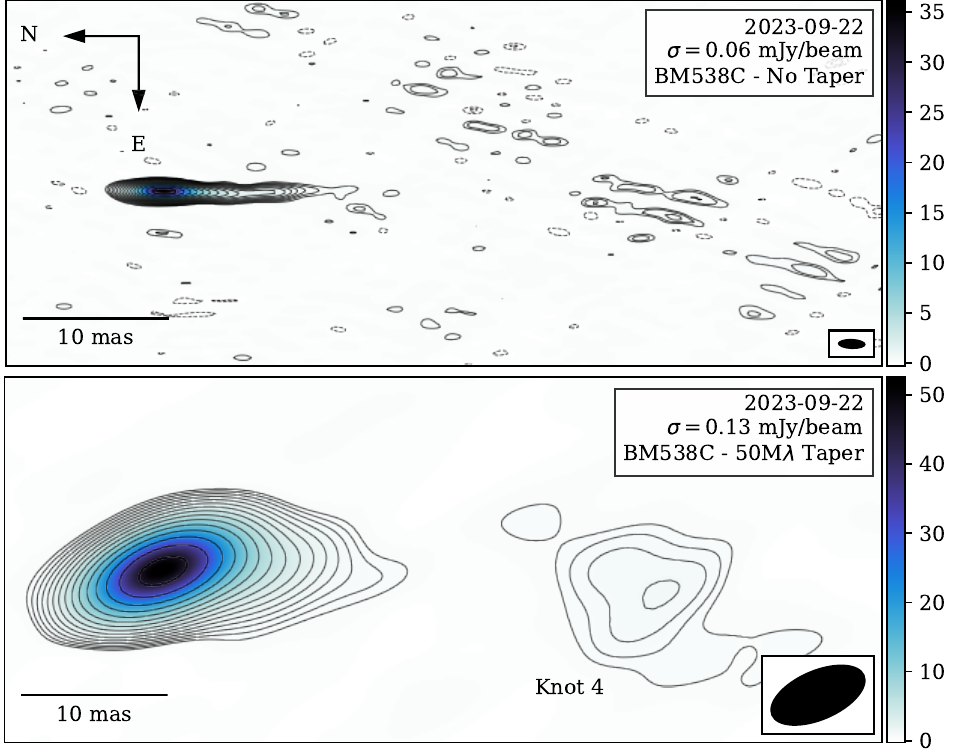}
                \caption{VLBA images of \sjfull\ on 2023 September 22 (MJD 60209). The top panel shows an image made from the full visibilities, while the bottom panel shows an image made using a Gaussian taper with 30\% power at \qty{50}{\mega\wavelength}. The colour scale shows the intensity in \unit{\mJy\per\beam}, and the contours are at $\pm\sigma\times\sqrt{2}^{n}$ for $n=3,4,5,...$ where $\sigma$ is the noise level of the image shown in the label. The images are rotated counter-clockwise by \qty{90}{\degree}, as shown by the compass, and the ellipse in the lower right corner shows the restoring beam of the images. The observation details are given in Table~\ref{tab:vlbi observation log}. The approaching extended jet of \sjfull\ was $\sim$\qty{10}{\mas} long and there was a rapidly-moving, smeared-out discrete jet knot to the south of the core, which we refer to as knot 4.}
                \label{fig:bm538c}
            \end{figure}

\begin{figure}
                \centering
                \includegraphics[width=\linewidth]{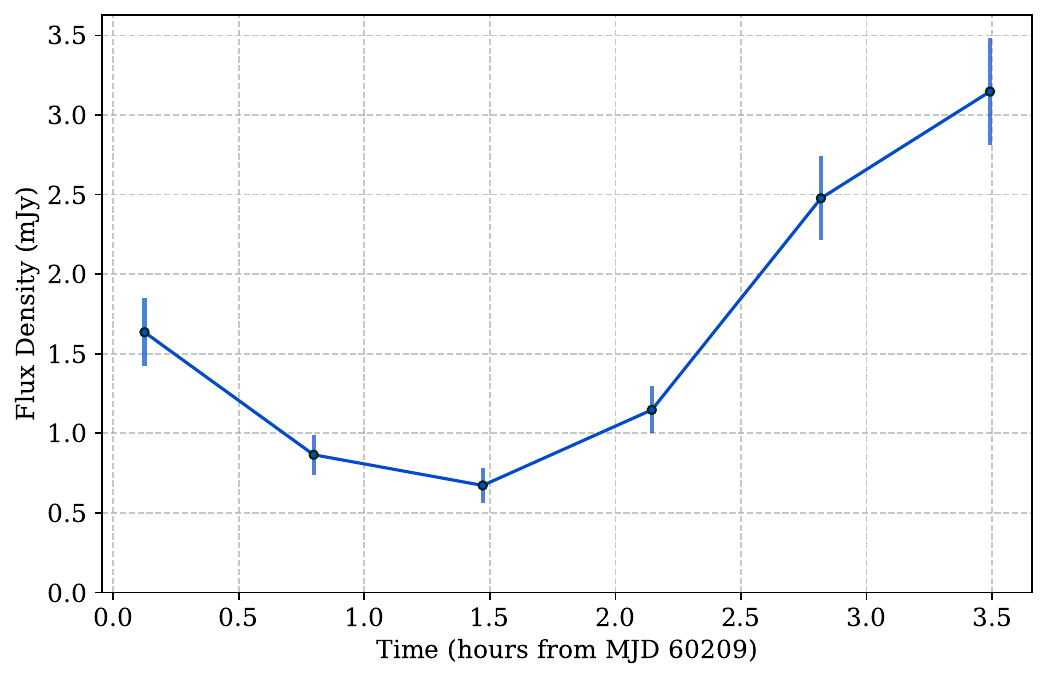}
                \caption{Light curves of knot 4 on 2023 September 22 (MJD 60209), derived from a model fit to observation BM538C. We used a piecewise linear flux density evolution, where we fit for the flux density at six fixed nodes (shown by the markers) and linearly interpolated between those nodes with five piecewise segments (shown with the solid lines). We added in a 10\% flux density calibration systematic error in quadrature with the statistical error of the fit.}
                \label{fig:bm538 C light curve}
            \end{figure}

            Figure~\ref{fig:bm538c} shows two images made from the flaring-period observation on 2023 September 22 (BM538C). The first image is made from the full visibilities in \aips\ without any visibility tapering, and the second in \difmap\ with a Gaussian taper of 30\% power at \qty{50}{\mega\wavelength}. These images show two key features; that in the three days since the previous observation \citep[BM538B;][]{2025ApJ...984L..53W}, the steady continuous jet had recovered from an extent of 2 to approximately \qty{10}{\mas}, and that there was a large, diffuse, downstream jet knot to the south of the continuous jet. The continuous jet had an integrated flux density of \qty{61\pm6}{\mJy}. Following the same approach as for the first VLBA observation \citep[BM538A;][]{2024ApJ...971L...9W}, we used the CLEAN components to $uv$-subtract the extended continuous jet from the phase-only self-calibrated observation, and used the residual visibilities to model the downstream knot. We fit the knot with a single circular Gaussian component with five piecewise linear flux density segments. We found that the jet knot was compact but moving rapidly, with a FWHM size of \qty{0.74\pm0.03}{\mas}, and a proper motion of $3.36_{+0.05}^{-0.04}$ \unit{\mas\per\hour}. This suggests that its diffuse appearance in the images is due to smearing as it moved rapidly during the observation, as opposed to it being a relatively stationary diffuse jet knot. Based on its position and proper motion, we cannot associate this jet knot with any previously ejected knots, and therefore we identify it as a new jet knot which we label `knot 4'. We show the light curve of knot 4 in Figure~\ref{fig:bm538 C light curve}, which shows that the flux density of the component was generally increasing over the course of the observation. \change{We discuss the nature of the flux density evolution of knot 4 (and of the jet knots detected later in the outburst) in Section~\ref{sec:variability discussion}.}
    
        \subsubsection{RM018}\label{sec:rm018}
\begin{figure}
                \centering
                \includegraphics[width=\linewidth]{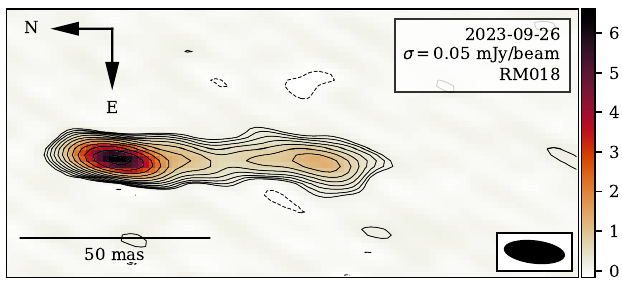}
                \caption{A 4.9-\unit{\GHz} EVN image of \sjfull\ on 2023 September 26. The colour scale shows the intensity in \unit{\mJy\per\beam}, and the contours are at $\pm\sigma\times\sqrt{2}^{n}$ for $n=3,4,5,...$ where $\sigma$ is the noise level of the image shown in the label. The image is rotated counter-clockwise by \qty{90}{\degree}, as shown by the compass, and the ellipse in the lower right corner shows the restoring beam of the image. The observation details are given in Table~\ref{tab:vlbi observation log}. This was the most extended image of the continuous jet from \sjfull\ in our VLBI campaign.}
                \label{fig:rm018}
            \end{figure}

            Figure~\ref{fig:rm018} shows an \aips\ image of our first EVN observation (RM018), taken during the flaring state, four days after BM538C. \change{Both EVN observations were taken at \qty{4.8}{\GHz}, and so they probed a slightly larger-scale structure than the 8.3-\unit{\GHz} VLBA and LBA observations.} The image shows an extended continuous jet with a downstream enhancement. It had an integrated flux density of \qty{9.55\pm0.98}{\mJy} and was $\sim$\qty{60}{\mas} in length. We computed the expected positions of all of the previously detected jet knots (0-4), but we found no evidence of emission from them in this observation. 

    \subsection{First State Transition}
        \subsubsection{RM019}\label{sec:rm019}
\begin{figure}
                \centering
                \includegraphics[width=\linewidth]{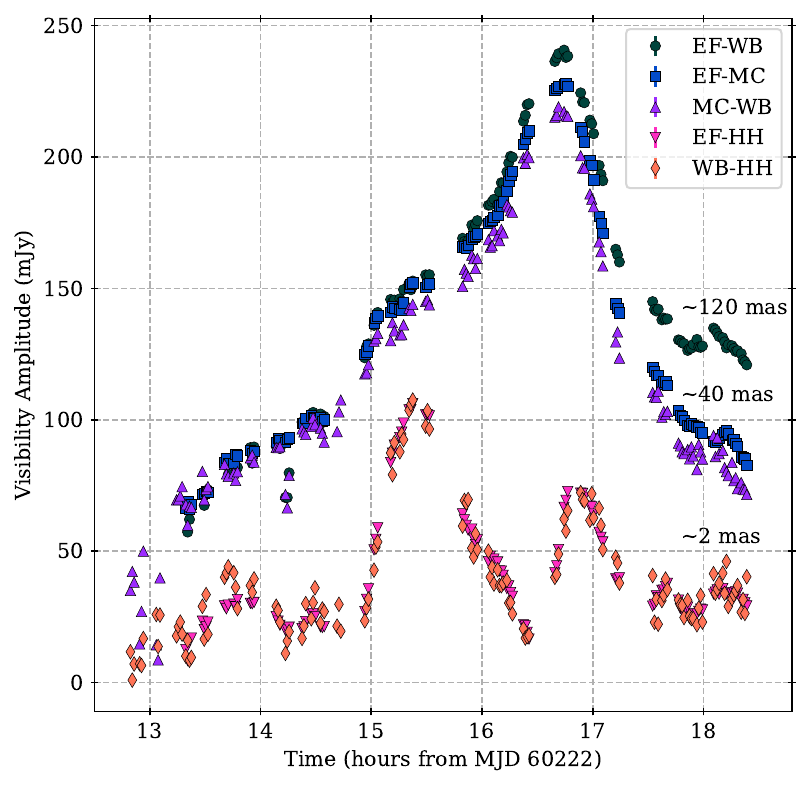}
                \caption{Visibility amplitudes from north-south baselines in the 4.9-\unit{\GHz} EVN observation of \sjfull\ on 2023 October 5 (RM019). The observation is centred at \qty{4.93}{\GHz}, unlike the VLBA and LBA observations (Table~\ref{tab:vlbi observation log}). The baselines between EF, WB, and MC are intra-European baselines, while baselines to HH are from Europe to South Africa. On the right side of the plot, we label the approximate north-south fringe spacing of the baselines (and therefore the size scale of the structure they probe). The visibility amplitudes show rapid variability on all baselines, including `beating' on the long baselines, likely due to the increasing separation of at least two distinct components.}
                \label{fig:rm019 amplitudes}
            \end{figure}

            Our second EVN observation (RM019) coincided with the first reported state transition of \sjfull\ on 2023 October 5 (MJD 60222). Analysis of this observation has previously been published in \citet{2025ApJ...987L..14C}. Figure~\ref{fig:rm019 amplitudes} shows the visibility amplitudes as a function of time from a selection of north-south oriented baselines. The black, blue, and purple markers are from `short' intra-European baselines, while the pink and orange markers show two baselines between Europe and Hartebeesthoek (HH), in South-Africa. We also show the rough size-scale probed by the baselines in the labels on the plot. A similar plot is shown in Figure~2 of \citet{2025ApJ...987L..14C}. Over the course of the $\sim5.5$ hour observation, \sjfull\ went from being $\sim50$ \unit{\mJy} to $\sim240$ \unit{\mJy}, and then back to $\sim125$ \unit{\mJy}. Towards the end of the observation, the shorter north-south baselines began to diverge, suggesting the source was becoming resolved in the north-south direction. The baselines to HH show more rapid amplitude variability with multiple peaks and troughs. \citet{2025ApJ...987L..14C} argued that this structure on the long baselines originates from the beating of the visibility amplitudes as two individual components move apart, which may be two distinct jet knots or a stationary core and a single jet knot. The $\sim2$ mas size-scale listed in Figure~\ref{fig:rm019 amplitudes} corresponds to the approximate fringe spacing of the Europe to HH baselines. When two point-source components are separated by an integer multiple of the fringe spacing, they will produce the maximum response from the baseline. Conversely, when they are separated by an integer multiple plus a half of the spacing, they will produce a minimum response from the baselines. \change{This is because the Fourier transform of a double point source is a cosine function with frequency proportional to the separation of the components \citep[see e.g.][for a demonstration of some simple visibility functions]{1999ASPC..180..335P}. Therefore, the growing separation of two components can produce beating of the visibility amplitudes as the location of the minima and maxima in the visibility response shift as a function of time \citep[see][for the first example of the measurement of the growing relative separation of a double source from visibility amplitudes]{1971Sci...173..225W}.} Based on the separation of the peaks on the long baselines in the second EVN observation, we estimate the apparent separation rate of the two components to be $\sim1-2$ \unit{\mas\per\hour}.

            \citet{2025ApJ...987L..14C} demonstrated this scenario with simulations of two jet knots (or a stationary core and a single jet knot) moving apart, and while they could replicate the beating of the long baselines, there are several features in the data that the simulations could not capture. The rapid divergence of the intra-European baselines is greater than can be explained by only the growing separation of two point source components, and necessitates intra-observation expansion of one or more of the jet components and/or for their motion to be more rapid and their associated ejection event (i.e the time at which they were at zero separation) to have been within the observation. The peaks on the baselines to HH, particularly the largest one at 15:20 UTC, are not time-symmetric in their rise and fall, which cannot be explained by only the growing separation of two components. Similarly, the peaks of the long baseline `flares' are different, and they are not proportional to the total flux density (as probed by the shortest baselines), suggesting that the two components have a changing flux density ratio and/or are changing in size during the observation.

            \begin{figure}
                \centering
                \includegraphics[width=\linewidth]{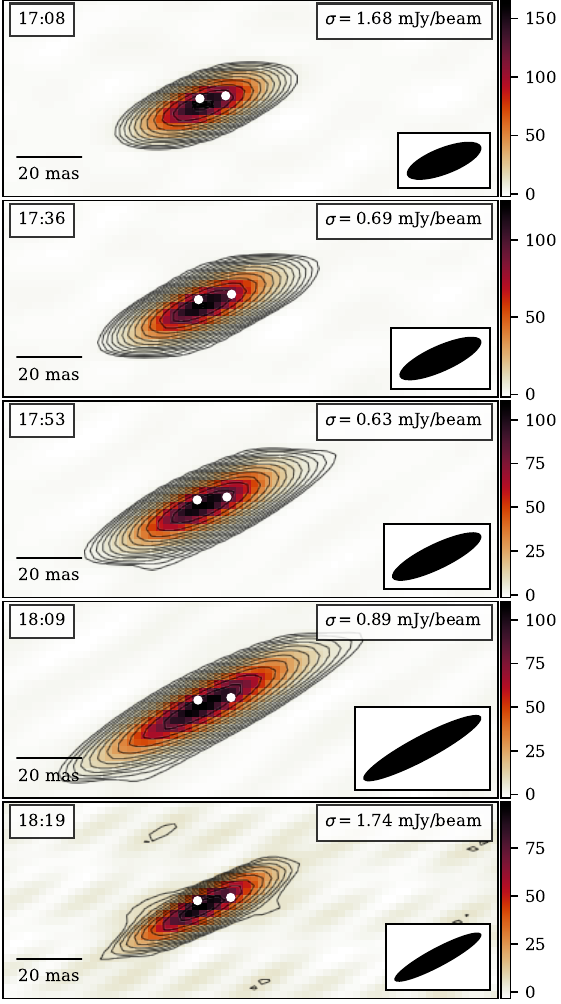}
                \caption{Series of \difmap\ images made with the intra-European baselines at the end of our second EVN observation of \sjfull\ (RM019), taken on 2023 October 5 (MJD 60222) at a central frequency of \qty{4.93}{\GHz}. Each image is made from $\sim20$ minutes of data, and is comprised of two point source components, which are moving apart at $\sim1-2$ \unit{\mas\per\hour}, as shown by the white markers. For each image, the colour scale shows the intensity in \unit{\mJy\per\beam}, and the contours are at $\pm\sigma\times\sqrt{2}^{n}$ for $n=3,4,5,...$ where $\sigma$ is the noise level of the images shown in the labels. The images are rotated counter-clockwise by \qty{90}{\degree}, and the ellipses show the restoring beams of the images.}
                \label{fig:rm019 montage}
            \end{figure}

            To explore these data further, we imaged the final $\sim1.5$ hours of the observation in \difmap. We split this period into five $\sim20$-minute segments and imaged them individually. To reduce the artefacts of sparse $uv$-coverage, we only imaged the intra-European baselines, giving us a nominal angular resolution of $~\sim20$ mas. We show these images in Figure~\ref{fig:rm019 montage}. We imaged in \difmap\, by fitting multiple static model components to the visibilities. We found that each segment was best represented by two point source components separated by less than the width of the beam. We show the location of these components as markers in the images in Figure~\ref{fig:rm019 montage}. At 17:08 UTC, the point sources were separated by $\sim$\qty{8}{\mas}, and by 18:19 UTC, they were separated by $\sim10$ mas, supporting our estimated $\sim1-2$ \unit{\mas\per\hour} of proper motion. Just as \citet{2025ApJ...987L..14C} described, it is not clear if these are two separate transient jet knots or a single approaching jet knot and the compact core. Due to poor external phase calibration and our reliance on phase self-calibration (which does not preserve absolute position information), we could not determine which of the two components, if either, were stationary. We did not have sufficient $uv$-coverage to fit resolved Gaussian components in these individual segments. Earlier in the observation, when imaging the Europe-only baselines, the multiple jet components were not sufficiently separated to be able to fit them with multiple distinct model components in individual segments. \citet{2025ApJ...987L..14C}, performed similar analysis of the Europe-only visibilities, however they only fit a single circular Gaussian component to each individual segment. While they found that towards the end of the observation, this component began to increase in size, by only using a single component they could not specifically capture the effect of motion in their imaging. They estimate that at the end of the observation, the `source' was expanding at $2.4\pm1$ \unit{\mas\per\hour}, however since they only fit the data with a single component, this `expansion rate' is a combination of the growing separation of the two components and their individual expansion rates. 

            To improve upon the broad constraints that we and \citet{2025ApJ...987L..14C} have drawn from inspecting the visibility amplitudes, and from imaging and model-fitting in \difmap, we used our time-dependent model fitting technique. While we were able to fit models to explain the structure of individual segments of the observation and subsets of the baselines, we were unable to explain the entire observation with a single consistent model.

            We found that by modelling the Europe-only baselines from the second half of the observation with two components, our best fit model consisted of two circular Gaussian jet components moving apart from each other, with the inferred ejection occurring around the time of the beginning of the rise of the brightest Europe-HH `flare' at $\sim$15:00 UTC. However, when we modelled the first half of the observation, and the observation in its entirety with the Europe-only baselines, the model instead favoured two circular Gaussian jet components moving apart from each other such that the ejection occurred prior to the beginning of the observation. We did not have sufficient $uv$-coverage to be able to constrain a model with three or more jet components (i.e. a core, a jet ejected prior to the beginning of the observation, and a jet ejected during the observation; or two approaching jet knots and two receding jet knots, etc.). As in \citet{2025ApJ...987L..14C}, because we relied on phase self-calibration, we cannot conclude if our two component models consist of a core and a single jet knot, or two jet knots moving apart from each other. 
            
            \change{We note that in all of our models with two components and in our \difmap\ images, the northern component was consistently brighter than the southern one. This disfavours an explanation where the northern component is a receding jet knot and the southern component is an approaching jet knot, since the approaching jet knot cannot be fainter than the receding jet knot (due to Doppler boosting). In Section~\ref{sec:speed and inclination} we discuss that the jet axis in \sjfull\ is relatively consistent in position angle and inclination throughout the outburst. Therefore, the northern component is highly unlikely to be approaching and the southern component receding. We suggest that either the northern component is the core and the southern component is an approaching jet knot, or they are both approaching jet knots with the southern knot moving faster than the northern knot (resulting in their growing separation). Without reliable external phase calibration, we cannot distinguish between these two explanations.}

            When we included the Europe-HH baselines in our modelling, we found that no physically reasonable model could adequately describe the rapid variability of the intercontinental baselines, although we note that due to the very sparse $uv$-coverage, we could still not reliably explore models containing more than two distinct jet components in the full data set. For this reason, we could not measure any reliable jet proper motions or infer any precise jet ejection dates from this observation, even with time-dependent visibility modelling. We can only agree with the conclusions of \citet{2025ApJ...987L..14C} that the rapid variability seen in this observations was most likely due to the motion and flux density variability of multiple discrete jet components (one of which may have been the core). Given the proximity of the components detected in imaging to the expected core position, we strongly suspect that one or more ejection events must have occurred just before and/or during this observation. We discuss some of the limitations of our time-dependent visibility modelling procedure, and in particular our failure to model either EVN observation, in Section~\ref{sec:modelling reliability}. 

        \subsubsection{BM538D, BM538E, BM538F, V456K, and BM538G}\label{sec:BM538D, BM538E, BM538F, V456K, and BM538G}
\begin{figure}
                \centering
                \includegraphics[width=\linewidth]{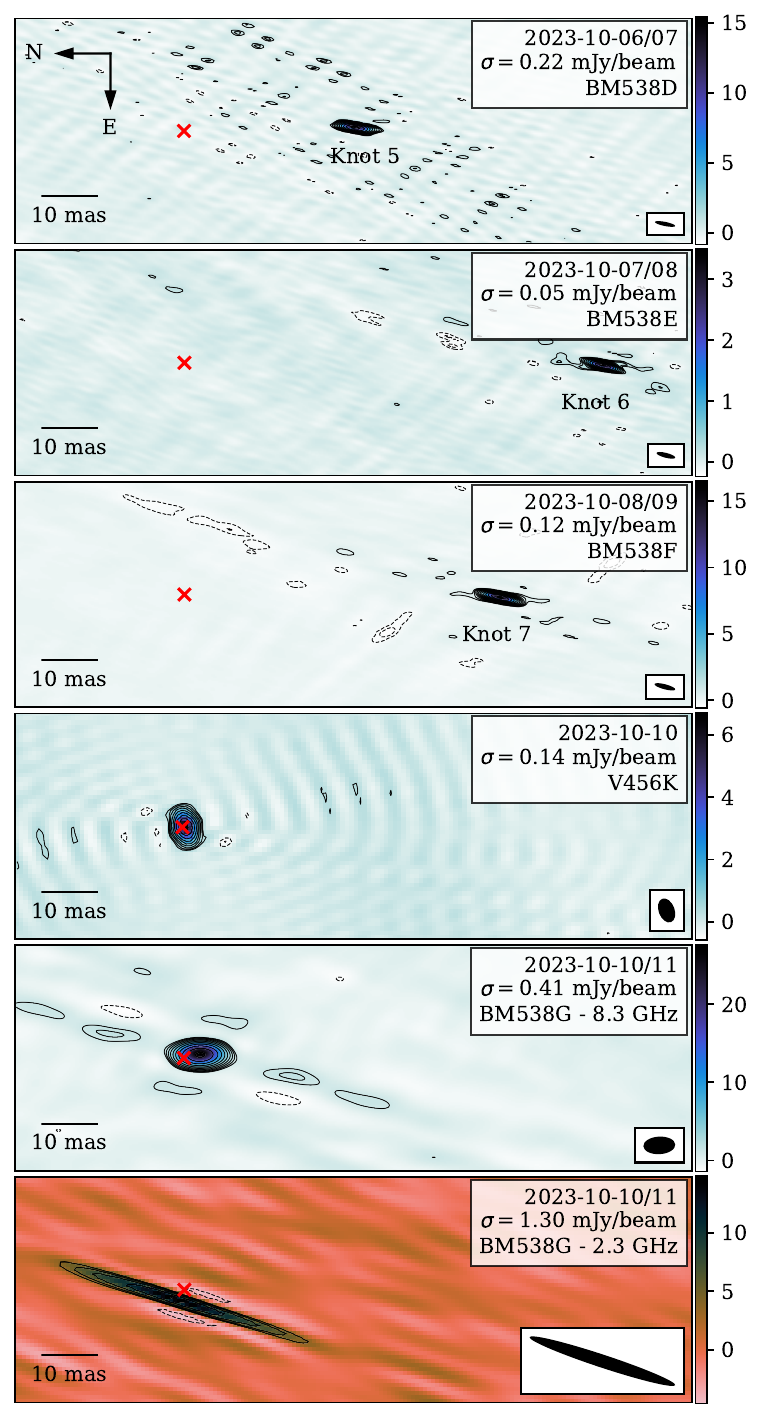}
                \caption{Images of \sjfull\ following the first reported state transition and associated radio flaring on 2023 October 5 (MJD 60222). For each image, the colour scale shows the intensity in \unit{\mJy\per\beam}, and the contours are at $\pm\sigma\times\sqrt{2}^{n}$ for $n=3,4,5,...$ where $\sigma$ is the noise level of the images shown in the labels. The images are rotated counter-clockwise by \qty{90}{\degree}, as shown by the compass, and the ellipses show the restoring beams of the images. The red crosses mark the location of the core in the first VLBA observation. The observation details are given in Table~\ref{tab:vlbi observation log}. The first three observations show three discrete transient jet knots (knots 5, 6, and 7), and by 2023 October 10 (MJD 60227) the compact core had re-established.}
                \label{fig:state transition montage}
            \end{figure}

            Figure~\ref{fig:state transition montage} shows images of the five observations following the reported 2023 October 5 hard-intermediate to soft-intermediate state transition (MJD 60222). For some of the observations we could only image a short section of the observation due to the rapid amplitude variability of the detected sources. The high noise level in each image in Figure~\ref{fig:state transition montage}, as well as the existence of imaging artefacts surrounding the detected jet components are further evidence of this rapid amplitude variability. In each observation shown in Figure~\ref{fig:state transition montage}, we detected a single jet component at varying separations from the inferred core location (marked with the black cross), and with different flux densities. Based on the location of the components, we infer that in each of the first three observations (BM538D, BM538E, and BM538F) we detected a discrete transient jet knot while the core was not detected. In the fourth and fifth observations (V456K and BM538G) we detected the re-established compact core, including at both 2.3 and 8.3 GHz in the fifth observation (BM538G). 

            To understand the nature of these jet knots, we modelled each observation individually. While we used several iterative rounds of model-fitting and self-calibration of the observations, we found that measuring the motion of the jet knots was difficult, since we relied entirely on the external phase calibration to fit the motion. This is because phase self-calibration does not preserve the absolute position of an isolated component in an observation, and so the proper motion we use to determine the self-calibration solutions (which is derived from the fit to initial external phase-calibrated data), is reinforced by the self-calibration. In other observations, we were able to `lock' the frame of reference through several rounds of self-calibration using the position of the stationary core as a fixed point of reference, which we were unable to do here. This issue was compounded by the fact that the VLBA observations during this period were plagued by poor weather and technical issues at a number of stations, resulting in poor external phase calibration.
    
            To measure the motion of the jet knots in the first three VLBA observations during this period, we had to fix the position angle of the jet motion, as some of our modelling results gave position angles of motion that were 5-20\unit{\degree} different from the jet axis (although the separations of the jet knots from the assumed core position were still aligned with the jet axis, $\sim180$\unit{\degree} East of North). To estimate the extra systematic uncertainty in the jet knot proper motion, we also fit the motion of the check source (J1724-2914). The observations of this source were calibrated alongside \sjfull\ using the same calibration solutions. It should be stationary within the observations, and so any apparent motion in the check source should reflect the systematic error in the external phase reference calibration. In each of the first three VLBA observations during this period, we found that the check source was apparently moving to the south with a proper motion of between 0.1 to 0.4\,\unit{\mas\per\hour}. For each observation, we added the measured proper motion of the check source as a systematic error in quadrature with the statistical error of the model fit to the intra-observation motions of the individual jet knots. For the observations where we did not detect the core, we also added in a systematic error of \qty{1}{\mas} to the separation of the jet knot from the inferred core location, in quadrature with the statistical uncertainty.
    
            Even after including the additional systematic uncertainty due to the phase referencing, we found that each of the three jet knots had inconsistent proper motions, and therefore we identified them as three separate, discrete transient jet knots, which we label `knot 5', `knot 6', and `knot 7'. We report the fitted proper motions of these jet knots in Table~\ref{tab:motion parameters}. We also note that in each observation during this period, we did not detect any emission at the expected locations of the jet knots seen in any other observation. 

            \begin{figure}
                \centering
                \includegraphics[width=\linewidth]{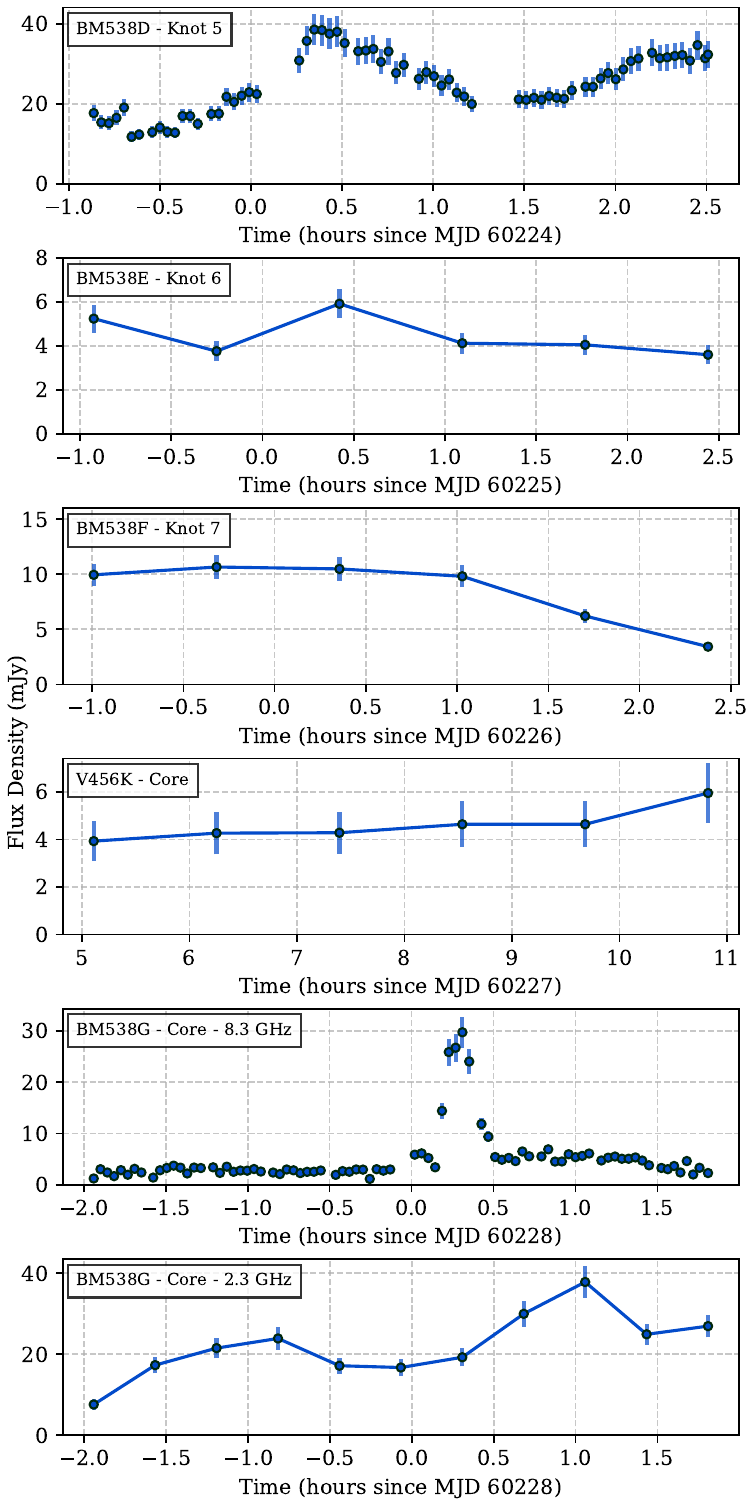}
                \caption{Light curves of the jet knots launched by \sjfull\ following the reported state transition on 2023 October 05 (MJD 60206). The y-axis shows the integrated flux density in mJy, measured by fitting for the flux density evolution of the jet components (transient knots and compact core) shown in Figure~\ref{fig:state transition montage}.  For observation BM538D and the 8.3 GHz band of BM538G, we fit the flux density in every individual scan, while in the others we linearly interpolated between the measured flux densities.}
                \label{fig:state transition light curve}
            \end{figure}
    
            For all of the jet components shown in Figure~\ref{fig:state transition montage}, including the compact core in V456K and BM538G, we used a constant-size elliptical Gaussian model, with a fixed position angle for knots 5, 6, and 7, to align with their fixed direction of motion. We summarised the fitted size parameters for these three jet knots and the compact core in Table~\ref{tab:size parameters}. For observation BM538D and the 8.4 GHz band of observation BM538G, we had enough signal to noise to individually fit the flux density of the source in every individual scan. For observations BM538E, BM538F, and V456K, we found that we could best represent the flux density evolution with five piecewise linear segments, and ten segments for the 2.3 GHz band of BM538G. In Figure~\ref{fig:state transition light curve} we show the intra-observation light curves of knots 5, 6 and 7, as well as the compact core in observations V456K and BM538G. 

            Knot 5 was the brightest and most variable jet knot that we observed and characterised during our campaign (excluding the unmodelled jet components in the second EVN observation). Within the $\sim$\qty{3}{\hour} observation in which it was detected, knot 5 increased from $\sim10$\,\unit{\mJy} to $\sim40$\,\unit{\mJy}, then faded to $\sim20$\,\unit{\mJy}, and then increased again to $\sim30$\,\unit{\mJy}. Knot 6 and 7 were fainter and more slowly evolving, both generally fading over the course of their respective observations. In the first observation of the re-established compact core (V456K), it was fairly constant in flux density. In the second observation of the core, it was also relatively constant at a similar flux density to the first observation, except for a remarkable, short-lived $\sim20$-minute long, $\sim30$ mJy flare at \qty{8.3}{\GHz}. There appears to be a similar flare in the \qty{2.3}{\GHz} light-curve, lasting $\sim2$ hours, and with the peak delayed by $\sim50$ minutes with respect to the \qty{8.3}{\GHz} light curve. We verified that this flare was intrinsic to the source by inspecting the intra-observation flux density evolution of the phase-referencing calibrator and the check source, and the amplitude calibration solutions applied in the external calibration process. We also observed this flare upon direct inspection of the visibility amplitudes of \sjfull. The flare may be related to the ejection of transient jets, however in our modelling we were unable to fit any additional model components to the observation, such as a rapidly evolving transient jet knot. \change{Assuming an apparent projected speed of $\beta_{\text{app}}\equiv\frac{v_\text{app}}{c}=1$}, a 50-minute delay corresponds to an angular separation of $\sim1$ mas, which is consistent with the size of the fitted elliptical component at 8.3 GHz (Table~\ref{tab:size parameters}). The rapid amplitude variability of knot 5 and of the core in observations BM538D and BM538G, respectively, explains the more prominent artefacts in the images of those observations shown in Figure~\ref{fig:state transition montage}.
    
    \subsection{Second State Transition}
        \subsubsection{BM538H and BM538I}
\begin{figure}
                \centering
                \includegraphics[width=\linewidth]{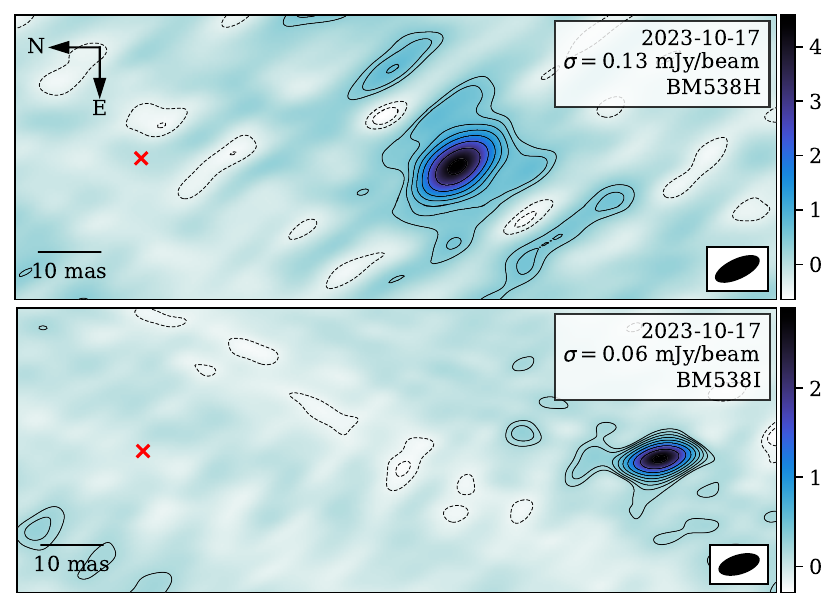}
                \caption{Images of \sjfull\ on 2023 October 17, following the second reported state transition. For each image, the colour scale shows the intensity in \unit{\mJy\per\beam}, and the contours are at $\pm\sigma\times\sqrt{2}^{n}$ for $n=3,4,5,...$ where $\sigma$ is the noise level of the images shown in the labels. The images are rotated counter-clockwise by \qty{90}{\degree}, as shown by the compass, and the ellipses show the restoring beams of the images. The red crosses mark the location of the core in the first VLBA observation. The observation details are given in Table~\ref{tab:vlbi observation log}. The observations show a large, resolved discrete jet knot travelling between the observations at $\sim$\qty{30}{\mas\per\day}, which we refer to as knot 8.}
                \label{fig:bm538I bm538H images}
            \end{figure}

            Figure~\ref{fig:bm538I bm538H images} shows \difmap\ images of the two observations taken after the second reported hard-intermediate to soft-intermediate state transition. The images both show a bright, resolved, southern knot south of the core, which had switched off since our previous observation. If they were the same jet knot, then they had an apparent proper motion of $\sim$\qty{1.4}{\mas\per\hour}. Despite very similar observing conditions, the first image (BM538H) has a higher rms noise level than the second observation (BM538I), and more prominent image artefacts surrounding the detected jet knot, suggesting either poorer amplitude calibration, or more rapid amplitude variability of the jet knot, or both. We first fit the motion of the jet knots in the two observations separately, but found that their motions and separations were relatively consistent with being the same jet knot. We therefore identified this single jet knot as `knot 8', and we jointly fit the two observations with a single ballistic circular Gaussian component. We allowed the component to have a different constant size in each observation. We also performed the fit allowing for the linear expansion of the Gaussian component within each observation, however we found no evidence of intra-observation expansion of the jet knot. We found that the jet knot was moving at \qty{1.418\pm0.003}{\mas\per\hour} at a position angle of $179.29_{-0.09}^{+0.08}$\unit{\degree} East of North. In the first observation, the FWHM size of the jet knot was \qty{6.34\pm0.06}{\mas}, and in the second observation it was \qty{4.41\pm0.09}{\mas}, showing that the component was apparently more compact in the second observation. This is consistent with the images of the observations shown in Figure~\ref{fig:bm538I bm538H images}. 
    
            \begin{figure}
                \centering
                \includegraphics[width=\linewidth]{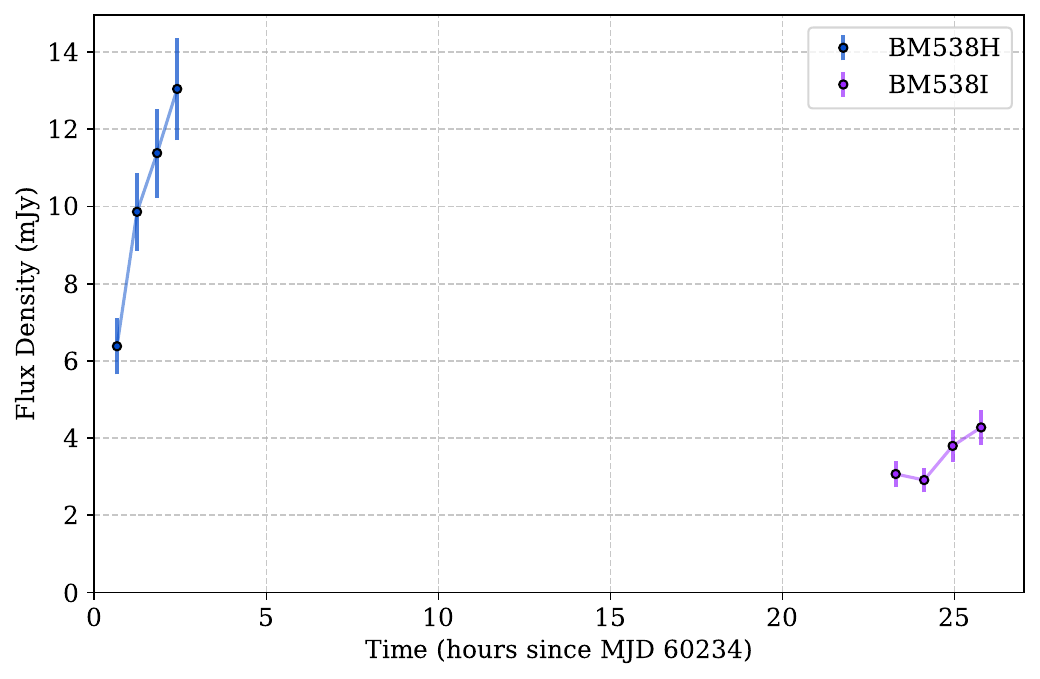}
                \caption{Light curves of knot 8 on 2023 October 17, derived from a joint model fit to observations BM538H and BM538I. We use a piecewise linear flux density evolution, where we fit for the flux density at four nodes (shown by the markers) and linearly interpolate between those nodes with three piecewise segments (shown with the solid lines). We add in a 10\% calibration systematic error in quadrature with the statistical error of the fit.}
                \label{fig:bm538I bm538H light curves}
            \end{figure}
            
            In each observation, we parameterised the light curve using three linear piecewise segments. We display these light curves in Figure~\ref{fig:bm538I bm538H light curves}, which show that during the first observation the jet knot was rapidly increasing in flux density, while it was much more steady in the second observation. This explains the higher noise level and more prominent image artefacts in the first image of this jet knot. These were our final observations of a transient jet knot launched during the the 2023-2024 outburst of \sjfull.

    \subsection{Reverse Transition}
        \subsubsection{BM538J}
\begin{figure}
                \centering
                \includegraphics[width=\linewidth]{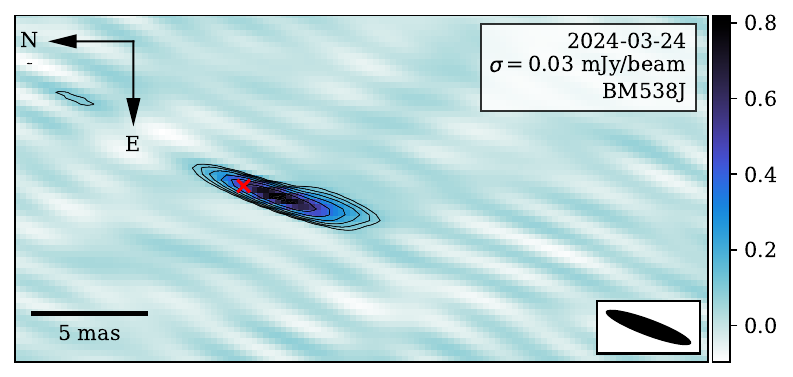}
                \caption{An image of \sjfull\ following the reverse transition in March 2024. The contours mark $\pm\sigma\times\sqrt{2}^n$ where $n=3,4,5,...$, and $\sigma$ is the rms noise shown in the upper right of the image. The ellipse in the lower right corner shows the restoring beam. The image has been rotated counter-clockwise by \qty{90}{\degree}. The red cross marks the position of the core in the first VLBA observation, showing the shift in the position of the core between August 2023 and March 2024.}
                \label{fig:bm538j}
            \end{figure}

            Figure~\ref{fig:bm538j} shows an image of \sjfull\ following the soft-to-hard reverse transition in March 2024. The continuous core jet had re-established, with an integrated flux density of \qty{0.76\pm0.09}{\mJy}. Despite the faintness of the source in this observation, the core jet was marginally resolved to the south ($\sim$\qty{1}{\mas}). The red cross marks the position of the core in the first VLBA observation, showing that from August 2023 to March 2024, there was a shift in the position of the core. 
            
            To explore this further, we measured the location of the core in each observation it was detected using the \aips\ task \texttt{JMFIT}, prior to performing any self-calibration. We also measured the position of the check source in the same way. The errors in the fit position of the core are the statistical errors reported by \texttt{JMFIT}, with a systematic error in the location of the core based on the VLBA estimated systematic errors of \citet{2006AA...452.1099P} added in quadrature. We conservatively assume the same systematic error for the LBA observations as for the much lower elevation VLBA observations. 
    
            \begin{figure*}
                \centering
                \includegraphics[width=0.8\linewidth]{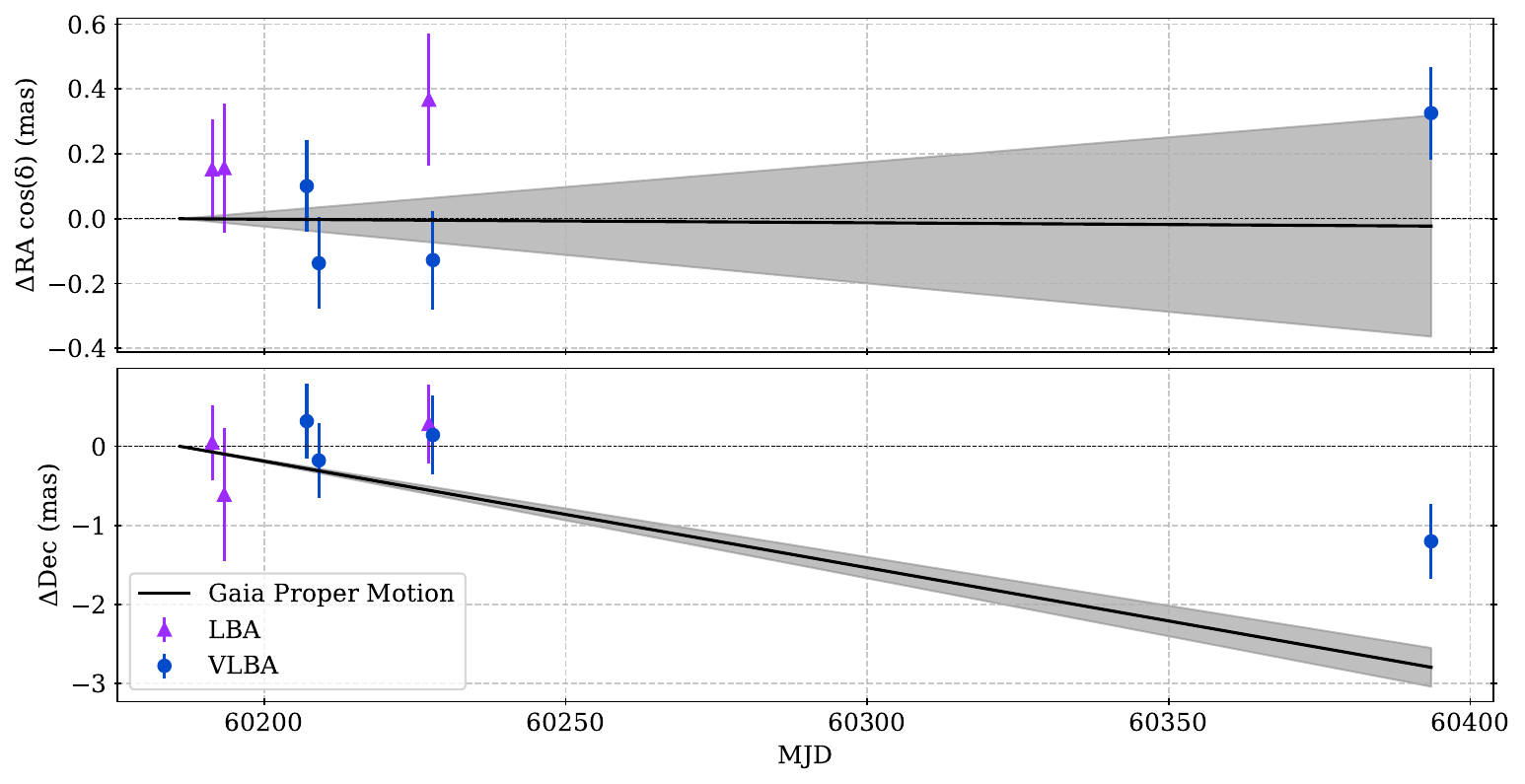}
                \caption{Shift in the position of the core of \sjfull\ from the first observations on 2023 August 30 (MJD 60186) with the VLBA (BM538A) and the LBA (V456H). The black line and shaded grey region show the proper motion of \sjfull\ measured by Gaia \citep[$\mu_\alpha\cos\delta=$\qty{-0.04\pm0.60}{\mas\per\year}, $\mu_\delta=$\qty{-4.92\pm0.43}{\mas\per\year};][]{2025A&A...693A.129M}. The deficit in the declination offset in the final VLBA epoch suggests that there may be a luminosity-dependent core-shift in \sjfull.}
                \label{fig:core astrometry}
            \end{figure*}
            
            For each epoch in which the core was detected, we computed the separation of the core position from its location on 2023 August 30. We do not include the EVN observations since they were taken at \qty{4.93}{\GHz}, and they had poor external phase calibration. To account for additional phase calibration errors, we measured the shift in the position of the check source in each observation which we subtracted from the \sjfull\ position shift. For the VLBA and LBA observations, we swapped the check source and phase reference source. However, we observed \sjfull\ with both instruments on 2023 August 30 (MJD 60186). For the VLBA observations, we computed the shift in the position in the core from the first VLBA observation (BM538A) and corrected for the change in the position of J1724-2914, and for the LBA observations we computed the shift in the position of the core from the first LBA observation (V456H), and corrected for the change in position of J1721-1628. 
            
            We plot the change in the location of the compact core in Figure~\ref{fig:core astrometry} alongside the proper motion of \sjfull\ measured by Gaia \citep{2025A&A...693A.129M}. This Figure shows that there was a shift in the position of the core upstream (i.e. in the direction of positive declination) between when it was detected in the bright hard-intermediate state at the beginning of the outburst, and in the faint hard-state following the reverse transition at the end of the outburst. There was also a marginal deficit in the offset of the core between the first observation and the observations taken on 2023 October 10/11 (MJD 60227/60228) when the core re-established following the first state transition. We discuss the possibility of a luminosity dependent core-shift in Section~\ref{sec:discussion core jet}. 

    \subsection{Summary of Transient Jet Parameters}\label{sec:Summary of transient jets}
        Across the full campaign, we detected and were able to fit the evolution of nine distinct jet knots. We were also able to fit a simple time-dependent model to the compact core in three separate observations. In the other observations, the core jet was either too extended to be fit with a simple model, or it was not detected. We also note that while the second EVN observation likely contained discrete transient jets, we were unable to fit their motion and evolution, due to the sparse $uv$-coverage and rapid source variability. We summarise the fit parameters of the nine distinct transient jet knots and the compact core in Tables~\ref{tab:motion parameters} and \ref{tab:size parameters}. Overall, we observed the jet knots to be travelling at a range of proper motions, and with varying sizes and flux densities, but at a relatively consistent position angle. Some of the jet knots were best fit with a constant size circular Gaussian, some with a linearly expanding circular Gaussian, and others with a constant-size elliptical Gaussian elongated along the jet axis.

        \begin{table*}
            \begin{center}
            \caption{Summary of the fit parameters for the position and motion of the time-varying jet knots. We report the median and 16th and 84th percentiles of the marginal posterior distributions for each parameter. The reference date was chosen to be close to the midpoint of each observation (or the midpoint between BM538H and BM538I), and the separation is the distance of the jet knot from the core location at the reference date. We jointly fit the motion of a single jet knot between observations BM538H and BM538I. For the observations in which we did not detect the core, we measure the separation using the best fit core location from the first VLBA observation, and we add in a systematic error of \qty{1}{\mas} in quadrature with the statistical error from the fits.}
            \label{tab:motion parameters}
            {\tablefont
            \begin{tabular}{@{\extracolsep{\fill}}lccccc}
                \toprule
                {Component} & {Observation} & {Reference Date} & {Proper Motion}          & {Position Angle}                       & {Separation} \\
                {}          & {}            & {(MJD)}          & {(\unit{\mas\per\hour})} & {(\unit{\degree} East of North)} & {(\unit{\mas})} \\
                \hline
                
                    Knot 0     & BM538A & 60186.083 & $0.66\pm0.04$          & $179.24\pm0.02$ & $66.89\pm0.04$           \\
                    Knot 1     & BM538B & 60207.083 & $0.62\pm0.06$          & $180.13\pm0.16$ & $8.54_{-0.08}^{+0.07}$   \\ 
                    Knot 2     & BM538B & 60207.083 & $1.1\pm0.2$            & $177.4\pm0.4$   & $17.4\pm0.2$             \\
                    Knot 3     & BM538B & 60207.083 & $3.18\pm0.10$          & $178.27\pm0.05$ & $64.66\pm0.11$           \\
                    Knot 4     & BM538C & 60209.083 & $3.36_{-0.05}^{+0.04}$ & $179.0\pm0.3$   & $33.33\pm0.05$           \\
                    Knot 5     & BM538D & 60224.042 & $1.2\pm0.4$            & $180^*$         & $29\pm1$             \\
                    Knot 6     & BM538E & 60225.042 & $3.5\pm0.4$            & $180^*$         & $76\pm1$             \\
                    Knot 7     & BM538F & 60226.042 & $1.9\pm0.12$           & $180^*$         & $58\pm1$             \\
                    \multirow{2}{*}{Knot 8} & BM538H & \multirow{2}{*}{60234.542} & \multirow{2}{*}{$1.418\pm0.003$} & \multirow{2}{*}{$179.29_{-0.09}^{+0.08}$} & \multirow{2}{*}{$67\pm1$} \\
                              & BM538I 
                \botrule
            \end{tabular}}
            \end{center}
            \begin{tabnote}
                {$^*$Fixed parameter.}\tnp
            \end{tabnote}
        \end{table*} 

        \begin{table*}
            \begin{center}
            \caption{Summary of the fit parameters for the sizes of the time-varying jet knots (and the compact core in three observations). We report the median and 16th and 84th percentiles of the marginal posterior distributions for each parameter. For elliptical Gaussian models, we forced the position angle of the jet motion to be the same as the position angle of the elliptical Gaussian major axis, and we list both the major and minor FWHM size. We jointly fit the motion of a single jet knot in observations BM538H and BM538I, although we allowed the size to be different in each observation. We were only able to fit a constant expansion model for knots 1, 2, and 3, and so we give their expansion rate, and FWHM size at the listed reference date (which are identical to the reference dates in Table~\ref{tab:motion parameters}).}
            \label{tab:size parameters}
            {\tablefont
            \begin{tabular}{@{\extracolsep{\fill}}lccccccc}
                \toprule
                {Component} & {Observation} & {Reference Date} & {Position Angle} & {FWHM Size} & {Expansion Rate} & {FWHM Major Size}   & {FWHM Minor Size}          \\
                {}          & {}            & {(MJD)}          & {(\unit{\degree} East of North)}   & {(\unit{\mas})} & {(\unit{\mas\per\hour})} & {(\unit{\mas})} & {(\unit{\mas})} \\
                \hline
                Knot 0     & BM538A & 60186.083 & - & $1.92\pm0.05$ & - \\
                Core       & BM538B & 60207.083 & $180.41\pm0.10$ & - & - & $2.015\pm0.012$ & $0.268\pm0.006$ \\
                Knot 1     & BM538B & 60207.083 & - & $0.89\pm0.05$ & $0.12\pm0.04$ & -& - \\ 
                Knot 2     & BM538B & 60207.083 & - &  $3.8\pm0.4$   &  $1.0\pm0.3$& -& -\\
                Knot 3     & BM538B & 60207.083 & - &  $3.78\pm0.18$ &  $0.25_{-0.17}^{+0.16}$ & -& -\\
                Knot 4     & BM538C & 60209.083 & - &  $0.74\pm0.03$ & - & -& - \\
                Knot 5     & BM538D & 60224.042 & $180^*$ & - & - & $2.17_{-0.015}^{+0.016}$&  $0.746\pm0.006$\\
                Knot 6     & BM538E & 60225.042 & $180^*$ & - & - & $4.58\pm0.16$ & $2.57_{-0.08}^{+0.07}$ \\
                Knot 7     & BM538F & 60226.042 & $180^*$ & - & - & $3.97\pm0.04$ & $1.473_{-0.019}^{+0.018}$ \\
                Core       & V456K  & 60227.042 & $180\pm2$ & - & - & $2.19_{-0.016}^{+0.015}$&  $0.6_{-0.4}^{+0.3}$ \\
                Core       & BM538G & 60228.333 & $177.5_{-1.5}^{+1.4}$ & - & - & $1.07\pm0.06$ & $0.29_{-0.03}^{+0.02}$ \\
                \multirow{2}{*}{Knot 8} & BM538H & \multirow{2}{*}{60234.542} &- & $6.34\pm0.06$ & -  & - & - \\
                           & BM538I &           &- & $4.41\pm0.09$   & - & - & - 
                \botrule
            \end{tabular}}
            \end{center}
            \begin{tabnote}
                {$^*$Fixed parameter.}\tnp
            \end{tabnote}
        \end{table*} 

        \subsubsection{Transient Jet Ejection Dates}    
            In \citet{2024ApJ...971L...9W}, we argued that knot 0, which we observed in the hard/hard-intermediate state, was more likely to have been some sort of downstream shock or jet-ISM interaction, as opposed to a discrete transient jet knot. For this reason, we did not compute the inferred ejection date of knot 0. For the other eight jet knots, we used their measured positions and proper motions (from Table~\ref{tab:motion parameters}) to compute their inferred ejection dates, assuming ballistic motion from launch to their observed separation from the core. We list these inferred ejection dates in Table~\ref{tab:ejection dates}, and note that the ejection dates of knots 1, 2, and 3 are as reported in \citet{2025ApJ...984L..53W}. While we suspect that one or more transient jets were ejected during or just prior to our second EVN observation (RM019), taken on 2023 October 5 (MJD 60222), we did not derive any conclusive ejection dates and therefore we do not list them here \citep[see also][]{2025ApJ...987L..14C}.

            \begin{table}
                \begin{center}
                \caption{The inferred ejection dates of the modelled jet knots, excluding knot 0. The ejection dates are calculated assuming ballistic motion, and using the proper motions and separations listed in Table~\ref{tab:motion parameters}.}
                \label{tab:ejection dates}
                {\tablefont
                \begin{tabular}{@{\extracolsep{\fill}}ll}
                    \toprule
                    {Jet Knot} & {Ejection Date} \\
                    {}          & {(MJD)} \\
                    \hline
                    Knot 1 & $60206.41_{-0.07}^{+0.06}$   \\
                    Knot 2 & $60206.36_{-0.17}^{+0.11}$   \\
                    Knot 3 & $60206.22\pm0.03$            \\
                    Knot 4 & $60208.670\pm0.006$          \\          
                    Knot 5 & $60223.0\pm0.3$              \\
                    Knot 6 & $60224.14\pm0.10$            \\
                    Knot 7 & $60224.77\pm0.08$            \\
                    Knot 8 & $60232.57\pm0.03$            
                    \botrule
                \end{tabular}}
                \end{center}
            \end{table} 
            
\section{Discussion}\label{sec:discussion}
    In 16 high angular resolution VLBI observations, we have studied the properties and evolution of the continuous and transient jets launched by the black-hole LMXB \sjfull\ during its 2023-2024 outburst. This was one of the most comprehensive high angular resolution monitoring campaigns of an LMXB in outburst. This unique set of observations allows us to probe the launching of both continuous and transient jets during multiple states of the outburst, and to investigate the properties of these jets and their coupling to the accretion flow, while controlling for variables such as black-hole mass, black-hole spin, spin-orbit misalignment, and distance. 
    
    In our images of \sjfull\, we detected numerous discrete transient jet knots, and with our time-dependent visibility modelling technique \citep[see e.g.][]{2023MNRAS.522...70W, 2025ApJ...984L..53W} we were able to precisely track the intra-observation motion of nine individual jet knots (see Section~\ref{sec:Summary of transient jets} for a summary of the transient jet parameters). We were also able to produce light curves of their intra-observation flux density evolution, revealing that some of the jet knots showed rapid ($\sim$hour timescale) intra-observation variability. This detailed study of the real-time evolution of these transient jet knots was only possible with time-dependent visibility modelling. These observations also reveal, for the first time, how the spatial extent of the continuous jet evolves as it quenches and reforms throughout a single outburst. 

    \subsection{Reliability of our Modelling Results}\label{sec:modelling reliability}
        In \citet{2023MNRAS.522...70W} we verified the time-dependent modelling technique with synthetic data. We found that we were able to reliably recover the motion, size, and flux density variability parameters of the synthetic data in the absence of calibration errors (i.e. with only thermal noise added). In \citet{2025ApJ...984L..53W} we first introduced a self-calibration procedure for analysing our second VLBA observation of \sjfull\ (BM538B), and we used it again here for the remaining observations of the outburst. 
        
        With phase-only self-calibration, the phase gains can be adjusted in such a way that the global position of the components in an image can be shifted, but their relative separation cannot. For observations where we detect and model multiple components, we can reliably measure their separation and relative motion. This allowed us to precisely measure the ejection dates of knots 1, 2, and 3, since we could fix the core position throughout the self-calibration and measure the relative motion of the jet knots \citep{2025ApJ...984L..53W}. For observations where we did not detect the core, we had to account for both the systematic uncertainty in the global position of the jet knots and the unseen core, and the systematic uncertainty in the intra-observation motion of the knots. As described in Section~\ref{sec:BM538D, BM538E, BM538F, V456K, and BM538G}, we estimated this systematic uncertainty in the core location and the intra-observation motion of the isolated jet knots by modelling the position and motion of the check source, which has a known location and should be static. For this reason, our best proper motion and ejection date constraints are from observations where we detect jet knots alongside the core. This is a dominant systematic error in determining the separation and proper motion (and therefore ejection dates) of isolated discrete jet knots.

        We were unable to model the intra-observation variability of our second EVN observation, primarily due to the sparse $uv$-coverage, and the complex structure and rapid intra-observation evolution of the source. One significant challenge with this observation was the large gap in $uv$-coverage between the intra-European baselines and the Europe to South Africa baselines. The rapid intra-observation variability on these drastically different size-scales made modelling impossible. This is further complicated by the difficulty of performing accurate amplitude calibration for arrays with a series of intercontinental baselines to a single distant antenna. The incorporation of future southern African telescopes, such as SKA-Mid, will be crucial for improving the long baseline $uv$-coverage of the EVN, and allowing for accurate amplitude calibration for these long baselines.

        One major limitation of our modelling approach is in accounting for the structure of the extended continuous jet. For some observations where the core jet was compact (e.g. BM538B, V456K, BM538G) we fit it with a small elliptical Gaussian component. However, this model is generally not well suited for the large, asymmetrical, extended structure of the core seen in many other observations (BM538A, V456H, V456I, V456J, BM538C, RM018, BM538J). Recently, \citet{2025ApJ...986L..35Z} fit an analytical model to the resolved spatial structure of the continuous jet from our first VLBA observation of \sjfull\ (BM538A). This model provides a more appropriate representation of the structure of the resolved continuous jet than an elliptical Gaussian \citep[see also][]{2006ApJ...636..316H, 2013MNRAS.432.1319P}. However, the computation of its 2-dimensional Fourier transform and its application to fitting the structure and variability of the extended continuous core in the visibility plane is beyond the scope of this work. 

    \subsection{Jet Kinematics and Properties}

        \subsubsection{Position Angle and Precession}\label{sec:position angle and precession}            
            In our observations of \sjfull, we did not see any large-scale variation in the position angle of the continuous jet. This is consistent with X-ray polarimetric observations, which showed that over the course of the outburst and following the reverse transition, there was no significant change in the X-ray polarisation position angle, which was parallel to the continuous jet position angle \citep{2024ApJ...968...76I, 2024A&A...686L..12P}. With time-dependent visibility modelling, we were able to precisely measure the position angle of multiple jet knots, and some of them differed significantly from each other, it was never by more than $\sim$\qty{2.5}{\degree} (see Table~\ref{tab:motion parameters}). The fit position angles were based on the location of the fitted jet knots, and did not include any uncertainty accounting for the lateral size of the components. The sizes of the jet knots implied opening angles similar to the scale of the changes in the fitted position angles of the jet knots (see Section~\ref{sec:sizes and opening angles}). We therefore conclude that we did not observe a significant change in the position angle of the transient jet knots throughout the outburst. 

            Large-scale changes in the position angle and/or inclination of the jet axis ($>$\qty{10}{\degree}) has been observed in a number of X-ray binaries, e.g. in SS 433 \citep{1979Natur.279..701A, 1981ApJ...246L.141H}, V404 Cygni \citep{2017MNRAS.469.3141T, 2019Natur.569..374M}, GRS 1915+105 \citep[][Motta et al. in prep.]{2025ApJ...986..108R}, and Circinus X-1 \citep{2019MNRAS.484.1672C, 2025MNRAS.544L..37C}. This is often explained as the result of large-scale precession of the jet axis, although the precession time-scales and underlying mechanisms appear to be different in each system. Recent observations have also seen small scale precession of the position angle of the base of the jet launched by M87$^*$ \citep{2023Natur.621..711C}, although we note that the mass-scaled physical size and timescale probed in cm-wavelength VLBI studies of X-ray binaries are much larger than for M87$^*$.

            Studies of the evolution of the X-ray timing properties in \sjfull\ suggested that the observed X-ray QPOs may originate from precession of the inner disk or the jet base \citep[see e.g.][]{2024ApJ...968..106Z, 2024MNRAS.529.4624Y, 2025ApJ...986....3L, 2024ApJ...973...59S, 2024MNRAS.529.4624Y, 2025MNRAS.543.1748M}, however we did not find evidence that this precession affected the large scale orientation of the continuous jet or the jet knots.

        \subsubsection{Intrinsic Speeds and Jet Inclination}\label{sec:speed and inclination}
                Given that the changes in the position angle of the transient jets were small, we assumed that they also had similar inclination angles. All of the jet knots were seen moving to the south of the core location, and in all observations where the extended continuous core jet was significantly resolved, it was most extended to the south. We therefore assume that the jet axis was inclined to the line of sight with the approaching jet direction to the south of the core. We did not detect any receding counterpart to any of the approaching jet knots, likely due to the receding jets being deboosted below the detection threshold of our observations (which we quantify in Section~\ref{sec:receding deboosting}). For an approaching discrete jet knot, the observed proper motion ($\mu_a$) is,

                \begin{equation}\label{eqn:approaching proper motion}
                    \mu_a = \frac{\beta\sin i}{1-\beta\cos i}\frac{c}{d},
                \end{equation}
                where $c$ is the speed of light, $i$ is the inclination of the jet axis, $d$ is the distance to the source, and $\beta c$ is the intrinsic jet speed \citep{1999ARA&A..37..409M}. With measurements of the proper motion of approaching and receding bipolar jet knots, we can uniquely solve for the inclination angle of the jet axis. However, with only a measurement of the proper motion of an approaching jet knot, we can only solve for possible combinations of the intrinsic speed $\beta$ and inclination angle $i$. We write the apparent speed of the approaching jet as $\beta_{\text{app},a}=\mu_ad/c$, and rearrange equation~\ref{eqn:approaching proper motion} for the intrinsic speed, giving,
        
                \begin{equation}\label{eqn:beta intrinsic vs i}
                    \beta = \frac{\beta_{\text{app},a}}{\sin i+\beta_{\text{app},a}\cos i}
                \end{equation}
    
                \begin{figure}
                    \centering
                    \includegraphics[width=\linewidth]{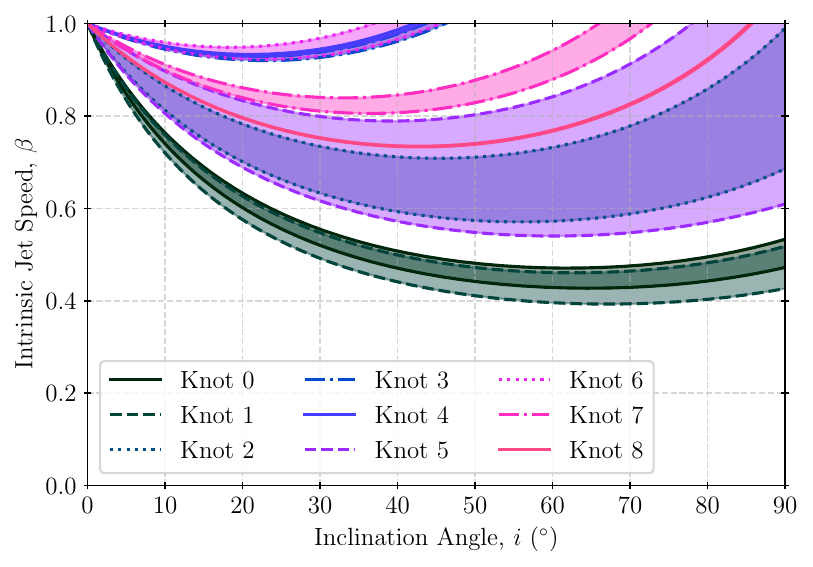}
                    \caption{Possible combinations of intrinsic jet speed ($\beta$) and inclination angle ($i$) for the nine jet knots that we modelled, at a fixed distance of \distanceNoErr. The shaded regions show the 16th and 84th percentiles as upper and lower bounds. The jet knots cannot be all be travelling at the same intrinsic speed if $i>0$, and the fastest travelling jet knots set an upper limit on the inclination when $\beta=1$. The upper limit on inclination set by the fastest jet knots is larger than what is shown in this plot once the uncertainty in the distance to \sjfull\ is included.}
                    \label{fig:intrinsic speed vs inclination}
                \end{figure}
        
                In Figure~\ref{fig:intrinsic speed vs inclination}, we plot equation~\ref{eqn:beta intrinsic vs i} for all nine jet knots listed in table~\ref{tab:motion parameters}. Similar curves have been plotted for knot 0 in \citet{2024ApJ...971L...9W} and knots 1, 2, and 3 in \citet{2025ApJ...984L..53W}, assuming a distance of \qty{3.7\pm0.3}{\kilo\parsec} \citep{2025A&A...693A.129M}. We did not incorporate any distance uncertainty in Figure~\ref{fig:intrinsic speed vs inclination}, since all the jet knots must be at the same distance. Incorporating the uncertainty in distance would increase the range of possible intrinsic speeds for each knot, but the uncertainties in the jet speeds at a given inclination will be correlated between jet knots, i.e. if the distance is smaller, then all the jet knots will be slower, and vice versa. Figure~\ref{fig:intrinsic speed vs inclination} shows that the nine jet knots must either have different intrinsic speeds, be at different inclination angles, or both. They can only all have the same inclination and jet speed at $i=0$. These curves also show that some of the jet knots have an upper limit on their possible inclination angles (when $\beta=1$). Only jet knots that are apparently superluminal (i.e. those with $\beta_{\text{app},a}>1$) have an upper limit on their possible inclination angle. By setting $\beta=1$ in equation~\ref{eqn:beta intrinsic vs i}, we derive an upper limit on the inclination angle of an apparently-superluminal jet knot of,
        
                \begin{equation}\label{eqn:max inclination angle}
                    i_\text{max} = 2\arctan\frac{1}{\beta_{\text{app},a}}.
                \end{equation}
    
                While we can use equation~\ref{eqn:beta intrinsic vs i} to \change{determine} the possible combinations of $\beta$ and $i$ for a given proper motion, each intrinsic speed is not equally as probable. Figure~\ref{fig:intrinsic speed vs inclination} shows that for a significant range of inclination angles, the intrinsic speed for many of the knots is close to their minimum value. The jet knots with the slowest proper motions can also only be travelling close to $\beta=1$ at very low inclination angles. \citet{2025A&A...693A.129M} measured a dynamical mass function for \sjfull\ of
                \begin{equation}
                    f(M_1) = \frac{M_1\sin^3i}{(1+M_2/M_1)^2} = 2.77\pm0.09 M_\odot,
                \end{equation}
                which imposes a limit on the minimum mass of the central black hole. For an orbital inclination less than $20$\unit{\degree}, the minimum black hole mass is $\geq70 M_\odot$, which is more than three times the mass of the black hole in Cygnus X-1 \citep[$21.2\pm2.2 M_\odot$;][]{2021Sci...371.1046M}, suggesting that a low inclination for \sjfull\ (i.e. $\lesssim20$\unit{\degree}) is highly unlikely. X-ray spectral and polarimetric modelling also suggests that the inner disk of \sjfull\ had a moderate inclination \citep[$\sim30-60$\unit{\degree};][]{2023ApJ...958L..16V, 2024ApJ...960L..17P, 2024ApJ...966L..35S}.

                \begin{figure}
                    \centering
                    \includegraphics[width=\linewidth]{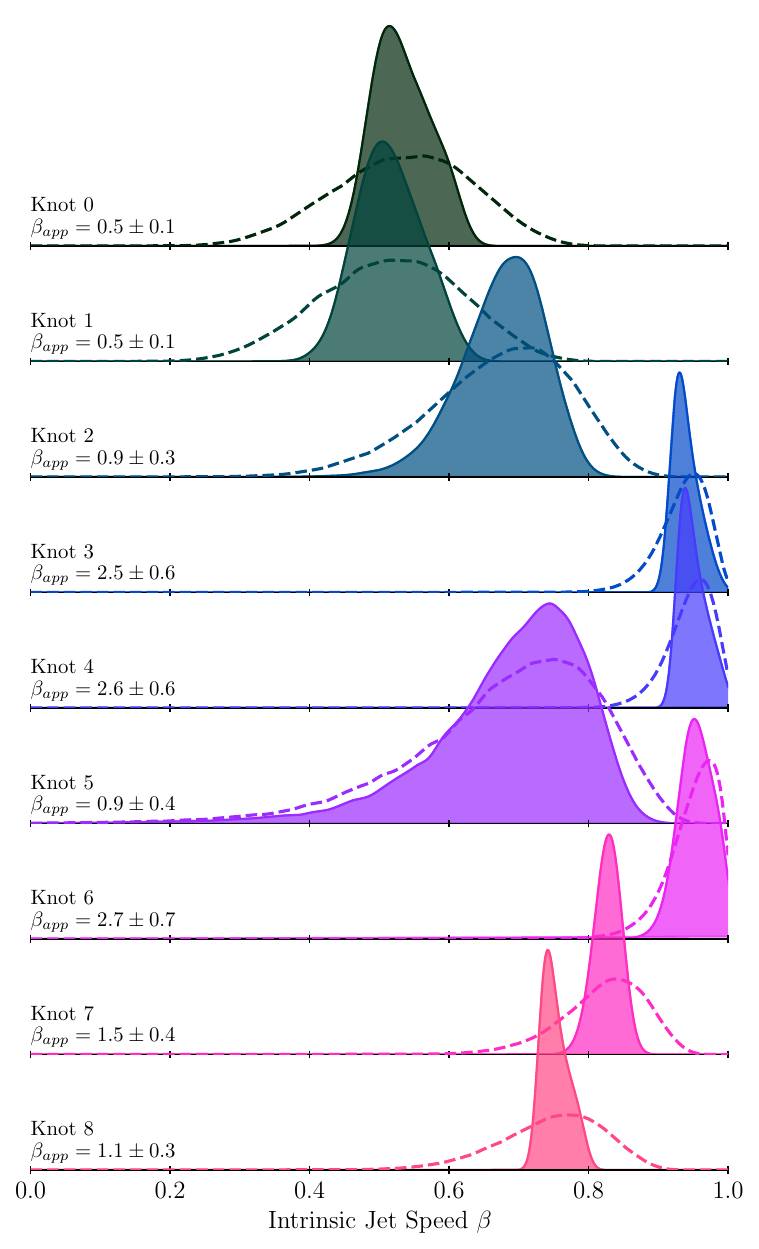}
                    \caption{Comparison of the marginal probability distributions of the intrinsic jet speeds of the nine modelled jet knots, derived from sampling an isotropic distribution of inclination angles and computing the jet speeds for each jet knot. The shaded distributions show the intrinsic speeds for a fixed distance of \distanceNoErr, and the dashed lines show the intrinsic speeds for a distance of \distanceBurridge\ \citep{2025ApJ...994..243B}. \change{The y-axes show probability density.} The labels show point estimates of the apparent speeds of the jet knots, and point estimates of their intrinsic speeds (derived from these posterior distributions) are given in Table~\ref{tab:intrinsic speed}.}
                    \label{fig:intrinsic motion marginal distributions}
                \end{figure}
    
                \begin{table}
                    \begin{center}
                    \caption{Measured intrinsic jet speeds and bulk Lorentz factors for the nine modelled jet knots assuming a distance of \distanceBurridge\ \citep{2025ApJ...994..243B}. The marginal posterior distributions for the intrinsic jet speeds are shown in Figures~\ref{fig:intrinsic motion marginal distributions}. The listed values are the medians and 16th and 84th percentiles of the marginal posterior distributions.}
                    \label{tab:intrinsic speed}
                    {\tablefont
                    \begin{tabular}{@{\extracolsep{\fill}}cccc}
                        Jet Knot & Intrinsic Speed, $\beta$ & $\Gamma$ & $\beta\Gamma$ \\
                        \hline\hline
                        Knot 0 & $0.54_{-0.10}^{+0.09}$   & $1.19_{-0.08}^{+0.11}$ & $0.64_{-0.16}^{+0.18}$ \\ 
                        Knot 1 & $0.52_{-0.11}^{+0.10}$   & $1.17_{-0.07}^{+0.10}$ & $0.61_{-0.15}^{+0.18}$ \\
                        Knot 2 & $0.68_{-0.11}^{+0.09}$   & $1.37_{-0.15}^{+0.20}$ & $0.9_{-0.2}^{+0.3}$ \\
                        Knot 3 & $0.94_{-0.04}^{+0.03}$   & $3.0_{-0.6}^{+1.1}$    & $2.8_{-0.7}^{+1.1}$ \\
                        Knot 4 & $0.95_{-0.04}^{+0.03}$   & $3.2_{-0.8}^{+1.6}$    & $3.1_{-0.8}^{+1.6}$ \\
                        Knot 5 & $0.70_{-0.15}^{+0.11}$   & $1.4_{-0.2}^{+0.3}$    & $1.0_{-0.3}^{+0.4}$ \\
                        Knot 6 & $0.96_{-0.04}^{+0.03}$   & $3.5_{-1.0}^{+2.5}$    & $3.3_{-1.0}^{+2.5}$ \\
                        Knot 7 & $0.83_{-0.07}^{+0.05}$   & $1.8_{-0.2}^{+0.3}$    & $1.5_{-0.3}^{+0.4}$ \\
                        Knot 8 & $0.75_{-0.08}^{+0.06}$   & $1.52_{-0.17}^{+0.22}$ & $1.2_{-0.2}^{+0.3}$ \\
                        \hline
                    \end{tabular}}
                    \end{center}
                \end{table}
                
\begin{figure}
                    \centering
                    \includegraphics[width=1\linewidth]{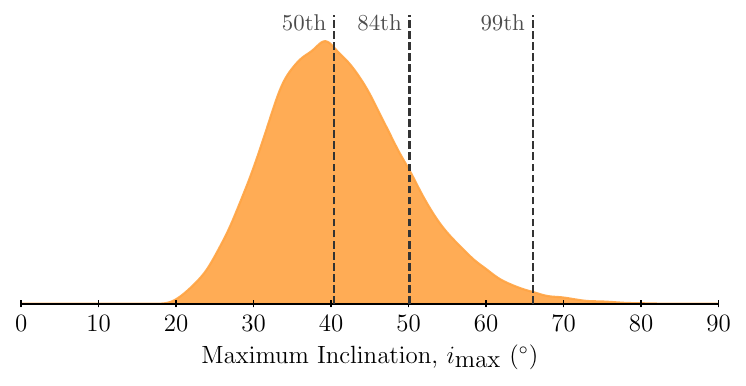}
                    \caption{Marginal posterior distribution of the maximum inclination angle of the jet axis computed from equation~\ref{eqn:max inclination angle} by sampling from the fastest jet knot proper motion posteriors and with a distance of \distanceBurridge\ \citep{2025ApJ...994..243B}. \change{The y-axis shows probability density.} The vertical lines show the 50th, 84th, and 99th percentiles of \qty{40}{\degree}, \qty{50}{\degree}, and \qty{66}{\degree}, respectively.}
                    \label{fig:max inclination}
                \end{figure}

                Based on these constraints, we computed a posterior probability distribution of the intrinsic speeds for each jet knot. To do this, we randomly sampled from the combined posterior distributions of the jet knot proper motions. For each sample of nine proper motions (one for each jet knot), we used the largest proper motion and equation~\ref{eqn:max inclination angle} to determine a maximum possible inclination angle. We computed this maximum value twice, once for a fixed distance of \distanceNoErr\ and once by sampling from the distance posterior distribution provided by \citet{2025ApJ...994..243B}. We then drew a random inclination angle from an isotropic distribution between a conservative lower limit of \qty{20}{\degree} and the maximum possible inclination angle for the sample of nine proper motions. With this inclination angle, we computed the intrinsic speed for each jet knot and created two joint posterior probability distributions of intrinsic proper motions and inclination angles (with and without the distance uncertainty included). We then integrated over the inclination angles to create marginal posterior probability distributions for the intrinsic speed of each jet knot. We show the intrinsic jet speed marginal distributions for both a fixed and sampled distance in Figure~\ref{fig:intrinsic motion marginal distributions}. The shaded distributions compare the intrinsic jet speeds of the jet knots when they are at the same distance, while the dashed distributions show our intrinsic jet speed constraints accounting for the distance uncertainty. In Table~\ref{tab:intrinsic speed}, we list the median and 16th and 84th percentiles of the intrinsic jet speed distributions for each jet knot, as well as their estimated \change{intrinsic} bulk Lorentz factor ($\Gamma=(1-\beta^2)^{-1/2}$), and the quantity $\beta\Gamma$, with the distance uncertainty incorporated. We also used our samples of $i_\text{max}$ (with the distance uncertainty incorporated) to create a marginal posterior distribution of the maximum inclination angle of the jet axis in \sjfull\ (Figure~\ref{fig:max inclination}), which had 50th, 84th, and 99th percentiles of \qty{40}{\degree}, \qty{50}{\degree}, and \qty{66}{\degree}, respectively. The dominant uncertainty in the maximum inclination angle is from the distance posterior. 

                Figure~\ref{fig:intrinsic motion marginal distributions} shows that \sjfull\ launched transient jet knots at a range of intrinsic speeds, with a clear distinction between the slower-moving mildly relativistic jet knots (knots 0, 1, 2, 5, 7, and 8), and the faster-moving highly-relativistic jet knots (knots 3, 4, and 6). Most previous high-resolution observing campaigns of transient jets from LMXBs have only detected a small number of ejecta. Studies of transient jet speeds have therefore relied on aggregating their properties over the population. However, this introduces systematic uncertainties due to the impact of intrinsic parameters that vary across the LMXB population, such as the nature and mass of the central compact object, the spin of the central compact object, any misalignment between the compact object spin axis and the binary orbital, and the existence of large-scale precession of the jet axis, etc. The varying intrinsic speeds of the jet knots launched by \sjfull\ imply that these parameters do not uniquely determine the ejection jet speed of transient jets. Furthermore, this implies that properties that vary throughout an LMXB outburst, such as the geometry and dynamics of the inner accretion flow and jet launching region, play a significant role in determining the ejection speed of the transient jet knots. 
                
                Recently, \citet{2025NatAs.tmp..198F} studied the speeds of a selection of mildly and highly relativistic transient jets launched by several neutron star and black hole X-ray binaries. They found evidence that the fastest moving jets are launched exclusively by black holes and always propagate along a fixed jet axis, and conversely, that transient jets launched along a precessing jet axis and/or by neutron stars are always slower. This suggests that the most relativistic transient jets propagate along the spin axis of the black hole, although they note that they found no evidence of a connection between the intrinsic jet speed and the reported black hole spin measurements. Our sample of both mildly relativistic and highly relativistic jet knots launched by the same LMXB during a single outburst allows us to further interrogate the connection between intrinsic jet speeds and jet precession. As discussed in Section~\ref{sec:position angle and precession}, the jet axis of \sjfull\ did not undergo any significant precession, and it launched both mildly and highly relativistic jet knots. This is similar to MAXI J1820+070, where both a mildly and a highly relativistic jet knot were both launched along a very similar position angle \citep{2021MNRAS.505.3393W}. In \sjfull, while two of the most relativistic jet knots, knot 3 ($\beta=0.94_{-0.04}^{+0.03}$, $\Gamma=3.0_{-0.6}^{+1.1}$) and knot 4 ($\beta=0.95_{-0.04}^{+0.03}$, $\Gamma=3.2_{-0.8}^{+1.6}$) marginally differed in position angle by \qty{0.7\pm0.3}{\degree}, this is not significant enough to rule out that their both of their jet axes were (or were not) aligned with the black hole spin axis. 

                \citet{2025ApJ...986L..35Z} fit the extended emission of the continuous jet resolved in the first VLBA observation of \sjfull, and found an intrinsic jet speed of \change{$\beta\sim0.3-0.4$}, assuming a distance of \qty{4}{\kilo\parsec}, an opening angle of \qty{1}{\degree}, and an inclination angle of \qty{45}{\degree}. At a larger inclination angle of \qty{65}{\degree}, the intrinsic jet speed was closer to \change{$\beta\sim0.6$}. The relationship between the assumed distance and the inferred jet speed of the continuous jet is not clear. This range of possible continuous jet intrinsic speeds is similar to our inferred intrinsic jet speeds for some of the slower-moving transient jet knots, however, the fastest-moving transient jet knots were significantly more relativistic than the continuous jet.
    
            \subsubsection{The Non-Detection of Receding Jet Knots}\label{sec:receding deboosting}
                We did not detect any receding counterparts to any of the approaching jet knots seen in our VLBI observations. The simplest explanation for this is that \sjfull\ launched intrinsically symmetric approaching and receding transient ejecta, but the receding counterparts were significantly de-boosted due to relativistic beaming. The ratio of the flux densities of two intrinsically symmetric approaching and receding jet knots is
                \begin{equation}\label{eqn:flux density ratio}
                    \frac{S_r}{S_a} = \left(\frac{1-\beta\cos i}{1+\beta\cos i}\right)^{3-\alpha},
                \end{equation}
                where $S_a$ and $S_r$ are the flux densities of the knots at equal angular separation from the core, and $\alpha$ is the spectral index of the jets \citep[$S_\nu\propto\nu^{\alpha}$;][]{1967MNRAS.136..123R, 1979Natur.277..182S, 1999ARA&A..37..409M}. Due to the light-travel time delay between the approaching and receding jet knots, this expression can not be used to estimate the expected flux density of a receding jet knot in an observation where an approaching jet knot was detected \citep{2004ApJ...603L..21M}. However, we can use our constraints on the intrinsic speeds and inclination angle to estimate the order of magnitude of the expected de-boosting of any receding counterparts to our nine jet knots, to comment on whether we should have expected any receding ejecta to be detected in our observations. 

                \begin{figure}
                    \centering
                    \includegraphics[width=\linewidth]{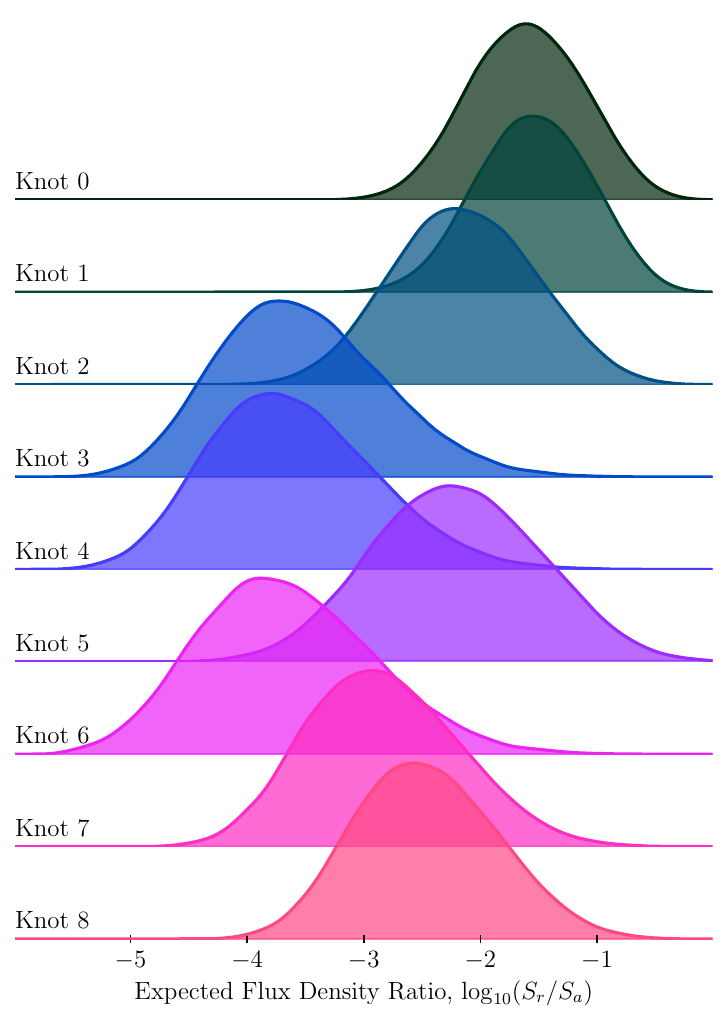}
                    \caption{Posterior distributions of the expected receding/approaching flux density ratios for the nine detected and modelled approaching jet knots, plotted logarithmically. The flux density ratios are calculated from equation~\ref{eqn:flux density ratio} using the sampled intrinsic jet speeds and inclination angles computed for Figure~\ref{fig:intrinsic motion marginal distributions}, with the distance uncertainty of \citet{2025ApJ...994..243B} included. \change{The y-axes show probability density.} The flux density ratio is only valid at equal separations from the core \citep{2004ApJ...603L..21M}, however we can use these constraints to estimate the expected order of magnitude of the flux densities of any undetected receding counterparts to the approaching jet knots.}
                    \label{fig:log doppler posteriors}
                \end{figure}
    
                With our sampled intrinsic speeds and inclination angles (with the distance uncertainty included), we computed the expected receding/approaching flux density ratio for each modelled jet knot, using equation~\ref{eqn:flux density ratio}, assuming a spectral index of $\alpha=-0.7$ \citep{2025ApJ...988..109H}. We \change{show} the distributions of the expected flux density ratio (on a log scale) for each jet knot in Figure~\ref{fig:log doppler posteriors}. Even for the slowest-moving jet knots (knots 0 and 1), the expected flux density of a receding counterpart at the same separation is 10-100 times fainter than the detected approaching knots. The receding counterparts to the fastest-moving jet knots may have been $10^3$-$10^5$ times fainter than their approaching counterparts. Given that we only detected jet knots with flux densities ranging from $\sim2-40$\,\unit{\mJy}, it is plausible that we did not detect any receding counterparts to the approaching jet knots entirely because of relativistic beaming due to their moderate inclination angle and moderate-to-high intrinsic speeds.

        \subsubsection{Flux Density Variability}\label{sec:variability discussion}
            In \citet{2024ApJ...971L...9W} and \citet{2025ApJ...984L..53W}, we fit the flux density evolution of knots 0, 1, 2, and 3, with simple linear flux density evolution. For the brighter jet knots detected later in the outburst (and the compact core in one observation), we constructed light curves of their intra-observation flux density evolution (Figures~\ref{fig:bm538 C light curve}, \ref{fig:state transition light curve}, and \ref{fig:bm538I bm538H light curves}). We also observed rapid changes in the visibility amplitudes in our second EVN observation immediately following the first reported state transition (see Figure~\ref{fig:rm019 amplitudes}), which is at least in part due to the rapid flaring of one or more transient jet knots. This short-time-scale variability was not captured in the $\sim$daily cadence short duration radio light curves presented in \citet{2025ApJ...988..109H}. The transient jet knots were relatively short-lived on VLBI scales, with only knot 8 detected in multiple observations, likely because they expanded and became resolved out relatively quickly, or that they rapidly faded below the detection threshold of our observations, or both. 
            
            Transient jet knots are often short-lived and rapidly variable on VLBI scales. The most extreme example is during the 2015 outburst of V404 Cygni, which repeatedly launched short-lived transient jets that showed rapid flux density variability from mm/sub-mm to radio wavelengths \citep{2019MNRAS.482.2950T, 2023MNRAS.518.1243F}. During a four hour period at the peak of V404 Cygni's 2015 outburst, \citet{2017MNRAS.469.3141T} modelled the multi-band light curves of a eight distinct bi-polar transient ejecta using the van der Laan synchrotron bubble model \citep{1966Natur.211.1131V}, where the individual knots adiabatically expand at a constant rate, with each knot having a single flare with an optically thick rise and an optically thin decay. During this same period, \citet{2019Natur.569..374M} resolved the motions of 12 individual ejecta within a single \qty{15}{\GHz} VLBI observation. Their intra-observation light curves, however, showed that the individual jet knots showed multiple peaks, complicating the synchrotron bubble model. This observation also showed that the core was always detected, was just as variable as the jet knots, and was the brightest component for a significant proportion of the observation. In \sjfull, we observed similar, complex, short-time-scale variability in our intra-observation light curves of multiple jet knots, particularly for knot 5 (top panel of Figure~\ref{fig:state transition light curve}). In \citet{2025ApJ...984L..53W} we discussed that the evolution of knots 1, 2, and 3 were not consistent with simple adiabatic expansion of a plasma bubble, and that interactions with the ISM or with previously ejected jet material may have caused their rapid variability and expansion. Differences in the internal properties of the jets, and the medium through which they propagated may explain the drastic differences in some of the properties of the jet knots, such as their flux density variability. 


        \subsubsection{Jet Sizes and Opening Angles}\label{sec:sizes and opening angles}
            Our images of \sjfull\ show the largest, most-resolved continuous jet ever seen in an X-ray binary. In our first VLBA observation (at \qty{8.4}{\GHz}) the continuous jet had a physical extent of $\sim\frac{230}{\sin i}\times(\frac{d}{5.5\,\text{kpc}})$ AU, or between $\sim200-820$ AU for a distance uncertainty of \distanceBurridge\ and an inclination range of \qty{20}{\degree}$-$\qty{66}{\degree}. In our first EVN observation (at \qty{4.9}{\GHz}), the continuous jet had a physical extent of $\sim\frac{400}{\sin i}\times(\frac{d}{5.5\,\text{kpc}})$ AU, or between $\sim340-1400$ AU for the same assumed distance and inclination range. These correspond to a mass-scaled size of $\sim(1-10)\times10^9\,r_g$ for a $10\,M_\odot$ black hole. We found that every single detected transient jet knot was intrinsically resolved. The sizes of the modelled jet knots are given in Table~\ref{tab:size parameters}. These correspond to physical sizes of the emission regions in the range of $\sim3-40$ AU.
            
            While jet transverse sizes and opening angles have been well studied in AGN \citep[e.g.][]{2017MNRAS.468.4992P}, they are generally poorly constrained in X-ray binaries, with only a handful of systems in which they have been measured \citep[see e.g.][]{2001MNRAS.327.1273S, 2006MNRAS.367.1432M, 2017MNRAS.472..141M,  2017MNRAS.469.3141T, 2020ApJ...895L..31E, 2021PASA...38...45C, 2021MNRAS.504.3862T, 2023MNRAS.522...70W, 2025ApJ...984L..53W}. This is in part due to difficulty resolving the jets transversely to the jet axis because of their smaller angular sizes than for nearby AGN, as well as sparse observing campaigns where the jet knots become resolved out between observations, so that their expansion cannot be tracked. Resolved transient jets at VLBI scales have only been observed in a handful of systems \citep[e.g.][]{2010MNRAS.409L..64Y, 2017MNRAS.468.2788R, 2020NatAs...4..697B, 2021PASA...38...45C}, and constraints on jet sizes are often confounded in high angular resolution observations by scatter broadening through the interstellar medium \citep[e.g.][]{1995Natur.374..141T, 1995Natur.375..464H, 2000ApJ...543..373D}. Given that the continuous jet was unresolved transverse to the jet axis at the full angular resolution of the VLBA, there is no evidence of scatter broadening in our observations, which is unsurprising given its high Galactic latitude of $\sim$\qty{10.9}{\degree}.

            \begin{figure}
                \centering
                \includegraphics[width=\linewidth]{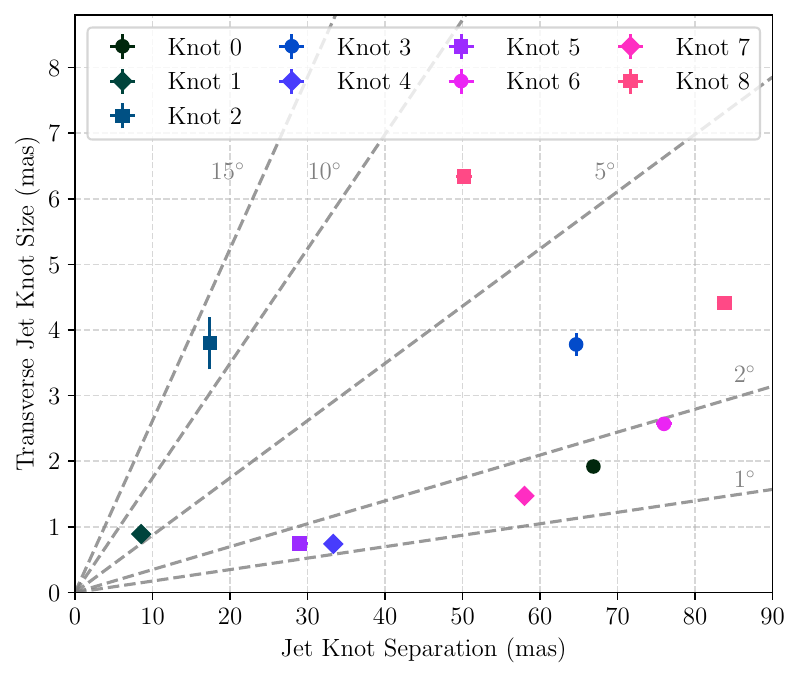}
                \caption{The FWHM size and separation of the nine detected and modelled transient jet knots. For the elliptical Gaussian jet knots, we plot the FWHM size perpendicular to the jet axis. The dashed lines show constant projected opening angles of \qty{1}{\degree}, \qty{2}{\degree}, \qty{5}{\degree}, \qty{10}{\degree}, and \qty{15}{\degree}. The jet knots show a range of projected opening angles which are all larger than the opening angle of the continuous resolved jet ($<1$\unit{\degree}).}
                \label{fig:knot size vs sep}
            \end{figure}
        
            From our first VLBA observation, where the continuous jet was most well resolved, we compute an upper limit on the projected opening angle of the continuous jet of $\lesssim$\qty{1}{\degree}. For the nine modelled transient jet knots, we plot the transverse jet knot FWHM size as a function of their separation in each observation where they were detected, in Figure~\ref{fig:knot size vs sep}. We also show lines of constant projected opening angle of \qty{1}{\degree}, \qty{2}{\degree}, \qty{5}{\degree}, \qty{10}{\degree}, and \qty{15}{\degree}. The jet knots had a range of projected opening angles between $\sim1-15$\unit{\degree}, which were all larger than the opening angle of the continuous jet. The difference in opening angles of the transient jet knots and the continuous jet may be related to the different confinement mechanisms of the phenomena \citep[e.g.][]{2023ApJ...954L..30S}, or different internal pressures of the jet material, which drive the rapid expansion of the transient jet knots. It may also be related to differences in the internal composition of the transient jets and the continuous jets \citep[e.g.][]{2024ApJ...967L...7Z}.

            While most jet knots were best fit with a circular Gaussian structure, some jet knots were found to be extended along the direction of the jet axis. In particular, the elongated jet knots were knots 5, 6, and 7, which were all detected following the first reported state transition. A VLBI observation of MAXI J1820+070 revealed an extended jet knot elongated in the direction of the jet axis \citep{2020NatAs...4..697B}. Of the two transient jet knots launched during its 2018 outburst, the elongated jet knot was moving more slowly and was brighter than the second, faster-moving jet knot. However, it was shown that the faster-moving jet knot was significantly deboosted due to its intrinsic speed and line-of-sight inclination \citep{2021MNRAS.505.3393W}. In our observations of \sjfull, the elongated jet knots were in general brighter than the jet knots launched earlier in the outburst, although they were neither clearly all faster or slower than the fainter and more circular jet knots. Due to the intra-observation variability of the jet knots and their short lifetimes at VLBI scales, it is difficult to compare their properties since the jet knots were not consistently detected at similar separations from the core. With that in mind, we did not find a strong relationship between jet knot speed, angular size, elongation, and flux density, although we note that any relationship between these properties will be significantly confounded by relativistic beaming effects, which are difficult to estimate without a precise measurement of the inclination angle.

            As discussed in Section~\ref{sec:variability discussion} and in \citet{2025ApJ...984L..53W}, the jet knots did not evolve following simple adiabatic expansion of a spherically symmetric synchrotron plasma cloud. Internal shocks driven by jet-ISM interactions or collisions between transient jets and previously ejected jet material may have been driving the detected emission, which would also impact the size and expansion of the jet knots. Their varying sizes may also be due to differences in internal composition and magnetic field pressure. Between the two observations in which it was detected, knot 8 faded and became smaller. This may be because the jet knot had become partially resolved out at VLBI scales, and thus we only detect the brightest, most compact part of the emitting region, or it could be because the emission is the result of internal shocks that create localised hot-spots that we detected and resolved. The complex evolution of these jet knots in our VLBI observations cannot be fully explained with simple models. Simulations of these ejecta may be necessary to fully understand their internal properties and the physical processes driving their emission and expansion \citep[e.g.][]{2025MNRAS.540.1084S}.

        \subsubsection{Luminosity-Dependent Core Shift}\label{sec:discussion core jet}
            By measuring the location of the core of the continuous jet at \qty{8.4}{\GHz}, and accounting for the predicted Gaia proper motion for \sjfull\ (see Figure~\ref{fig:core astrometry}, we saw a clear shift in its position between August 2023 (MJD 60186) and March 2024 (MJD 60393) in the direction of positive declination (i.e. antiparallel to the approaching jet direction). For a distance of \distanceBurridge, the expected shift in the core position between the first and final VLBA observation due to astrometric parallax is $\sim$\qty{+0.2}{\mas} in right ascension. While this is comparable with our measured shift in the position of the core in R.A., it is also comparable to the uncertainty in the Gaia proper motion in the $\sim$6 month gap between our observations. Without observations of the core spanning multiple years, we cannot constrain the parallax of \sjfull. The expected parallax shift of \sjfull\ in declination is much smaller than in right ascension, of order $\sim$\qty{0.02}{\mas}, which is significantly less than the observed shift in the core position in declination between August 2023 and March 2024. There was also a marginal shift in the position of the core (with respect to the Gaia proper motion) in the same direction (antiparallel to the approaching jet direction) after the continuous jet re-established, following the first state transition (MJD 60227). 
            
            The upstream core shift was more significant when there was a greater difference in the continuous jet flux density, suggesting a luminosity-dependent core shift, where the location of the approaching jet photosphere is closer to the central compact object when the continuous jet was fainter and less physically extended. Using the model of \citet{1979ApJ...232...34B}, \citet{2006ApJ...636..316H} derived an analytical expression for the jet luminosity ($L_\nu$) in Cyg X-1, finding that the photospheric radius at a given frequency ($z_\nu$) should scale with luminosity as $z_0\propto L_\nu^\frac{8}{17}$, for a constant jet speed and opening angle, and assuming that due to Doppler boosting the emission is entirely dominated by the approaching jet. \citet{2023MNRAS.525.4426P} found evidence of a luminosity dependent scaling in size of the continuous jet in MAXI J1820+070, based on upper limits on the shift in position of the jet photosphere from astrometric observations in the hard-state compared to estimates of the photospheric distance in radio timing observations when the radio jet was much more luminous \citep{2021MNRAS.504.3862T}. This is in agreement with our observations, which shows that at higher luminosities the peak of the jet was further downstream in \sjfull. We note that for a highly-resolved, apparently-asymmetric continuous jet (like we see in \sjfull), the location of the core (the peak of the continuous jet) will not be coincident with the location of the photosphere, since during the imaging process, we convolve the image with a 2D-Gaussian function (the so called restoring beam). The location of the peak will be biased downstream along the direction of the approaching jet as a result of the convolution. This effect will be more significant when the core is more extended.
            
            \citet{2025ApJ...986L..35Z} analysed the properties of the continuous jet in \sjfull\ by convolving the analytic jet profile with the Gaussian restoring beam, and then fitting it to the jet profile from our first VLBA observation, while also accounting for the pixel-to-pixel covariance in the measured jet profile. Similar analysis of the subsequent continuous jet images presented here may give more insight into the evolving continuous jet properties, and the apparent shift in the jet core position as a function of luminosity, however that is beyond the scope of this paper. 

    \subsection{Accretion/Ejection Coupling}
\begin{figure*}
            \centering
            \includegraphics[width=\linewidth]{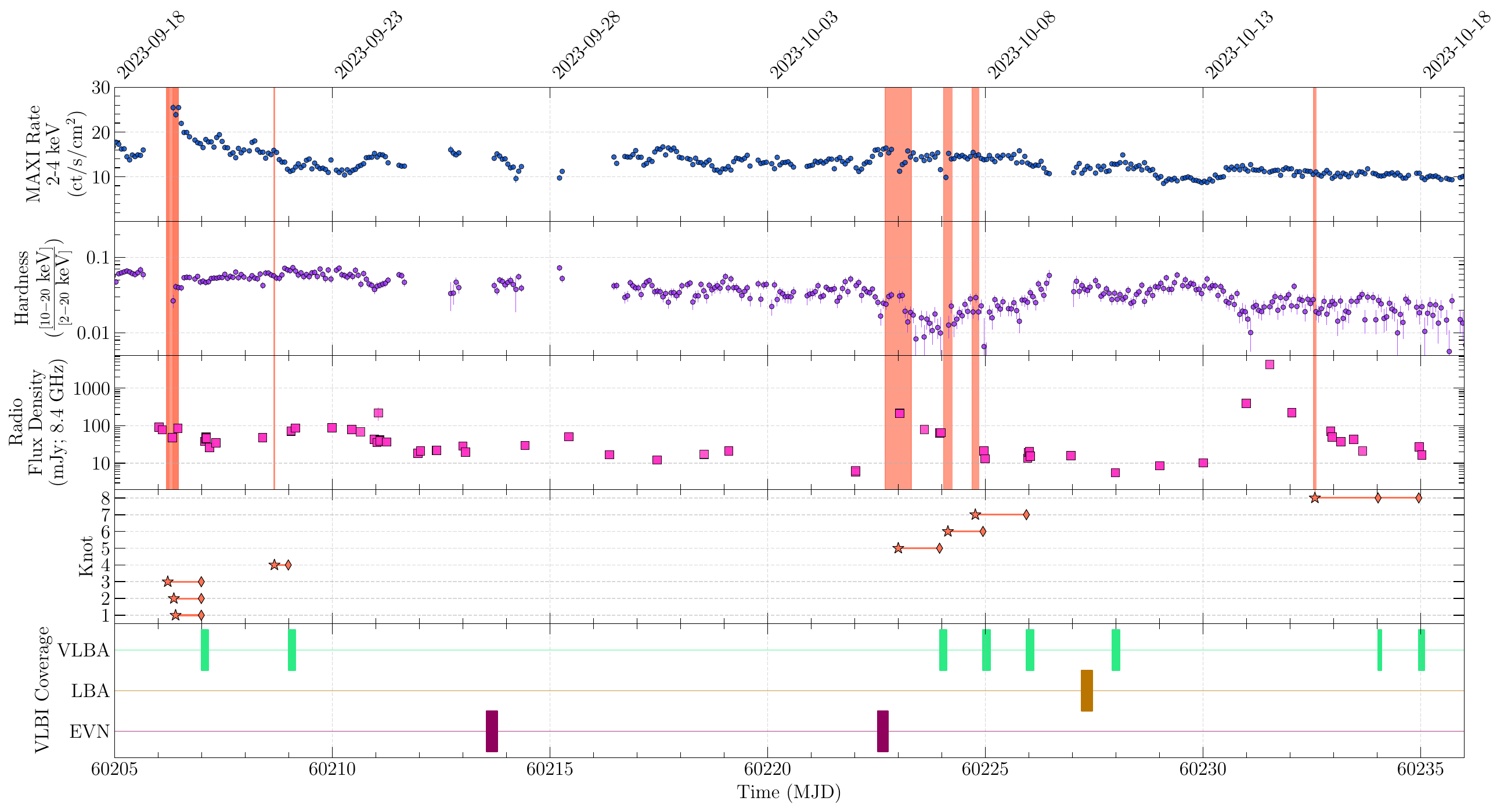}
            \caption{The ejection dates of the eight jet knots given in Table~\ref{tab:ejection dates} compared to the X-ray and radio coverage of the outburst. The first three panels and the bottom panel are the same as in Figure~\ref{fig:MAXI Lightcurve Observation Summary}, excluding the period of the initial hard and hard-intermediate-states. The shaded vertical lines show the uncertainty windows for our inferred ejection dates. In the fourth panel, the stars show the inferred ejection dates of the jet knots, and the diamonds show the observations in which the jet knots were detected. Knot 8 was the only transient jet knot detected in multiple observations. The other jet knots were only detected in observations within a day of their ejection. We note that there was strong evidence that transient jet ejection also occurred during or just prior to our second EVN observation on MJD 60222, which we did not plot here since we did not constrain a precise ejection date.}
            \label{fig:ejection date lightcurve}
        \end{figure*}

        In Figure~\ref{fig:ejection date lightcurve} we compare the timing of our observations, the measured ejection dates of the transient jet knots, and the overall X-ray and radio coverage shown in Figure~\ref{fig:MAXI Lightcurve Observation Summary}. This reveals a complex picture of accretion/ejection connection in the 2023-2024 outburst of \sjfull. Figure~\ref{fig:ejection date lightcurve} also shows the inferred ejection dates of all of the jet knots were all within $\sim1$ day of our observations, and only knot 8 was detected across multiple observations. There were also clearly large gaps in our coverage where transient jets may have been ejected, that we were not sensitive to. It also shows that while the short-duration, $\sim$daily cadence, low-angular-resolution radio light curves were able to broadly identify periods of transient jet ejection, they can not be used to estimate specific transient jet ejection dates, nor can they be used to track the $\sim$hour time-scale variability of the transient jets. While apparently-superluminal transient jets launched by \sjfull\ were resolved out to arcsecond scales (A. Hughes, priv. comm., 2025), it is not clear which of the many ejecta seen in our observations (if any) eventually travelled out to these large scales. 
        
        \subsubsection{Jet Evolution}

            Throughout its outburst, we observed how the spatial extent of the continuous jet of \sjfull\ evolved as it quenched and reformed. However, we observed and imaged the continuous jet with multiple instruments, at multiple frequencies, and with different sensitivities and restoring beams. We were able to fit the structure of the core with a compact elliptical Gaussian in only two observations (BM538B and BM538G). As a result, it is difficult to precisely and consistently measure and compare the extent of the continuous jet and it's integrated flux density between observations. As discussed in Section~\ref{sec:discussion core jet}, novel approaches to analysing the extended jet structure, such as the one used in \citet{2025ApJ...986L..35Z}, will be necessary to robustly constrain and compare the properties of the continuous jet throughout the outburst in our high-resolution observations. 
            
            We observed the resolved continuous jet of \sjfull\ several times throughout the outburst following our initial detections in the bright hard/hard-intermediate state. In the first VLBA observation in the hard/hard-intermediate state (MJD 60186), the continuous jet was highly extended ($\sim$\qty{30}{\mas}) and faded and contracted slightly in the subsequent LBA observations as the X-ray spectrum softened gradually and \sjfull\ progressed deeper into the hard-intermediate state. 
    
            Following the brightest X-ray flare in the flaring state (MJD 60206), we detected and tracked knots 1, 2, and 3, finding that their ejection dates were consistent with the peak of this flare, and a sudden change in the spectral and timing properties of the inner accretion flow \citep{2025ApJ...984L..53W}. Based on their analysis of X-ray timing observations,  \citet{2026A&A...706A.208J} suggested that during the bright X-ray flare on MJD 60206, \sjfull\ may have briefly transitioned into the soft-intermediate state. Following this flare, the continuous jet had significantly faded and become much more compact ($\sim$\qty{2}{\mas}), suggesting a partial (or possibly complete) quenching of the continuous jet during this brief transition, which is supported by the overall radio light curves, which dipped and became optically thin \citep[Figure~\ref{fig:ejection date lightcurve}; see also Figure~4 of][]{2025ApJ...988..109H}. 
            
            In our subsequent observation (Figure~\ref{fig:bm538c}, which was taken two days later during the flaring period (MJD 60209), the continuous core jet had become extended again ($\sim$\qty{10}{\mas}), and we detected and tracked knot 4. The inferred ejection date of knot 4 occurred around the time of a small dip in X-ray hardness and a change in the 2-4 keV X-ray intensity, although this was significantly less drastic than the flare on MJD 60206 (which corresponded to the ejection of knots 1, 2, and 3). Detailed analysis of X-ray timing observations around the ejection date of knot 4 may provide more insight into the evolution of the accretion flow during this period. At the time of this observation, the overall radio light curve had recovered back to a similar flux density as before the bright X-ray flare, and become optically thick again. Four days later, in our first EVN observation (MJD 60213; Figure~\ref{fig:rm018}), the continuous jet was highly extended ($\sim$\qty{30}{\mas}), despite being relatively faint. This was our only resolved image of the continuous jet at \qty{4.9}{\GHz}. The large extent and low integrated flux density of the continuous jet in our first EVN observation implies that much of the extended emission was diffuse and optically thin, and that the continuous jet was likely more extended than was seen in our \qty{8.4}{\GHz} observations. This supports our explanation for knot 0, which was that it was likely an internal shock in the more diffuse extended continuous jet material, or the result of an interaction between the diffuse extended continuous jet emission and the interstellar medium. Our EVN image of the continuous jet showed a bright, downstream feature at the tip of the continuous jet, which also could be the result of internal shocks or a jet-ISM interaction. This may also explain the origin of knot 4, which, similar to knot 0, was detected in an observation where the continuous jet was highly extended, although its highly relativistic speed is inconsistent with estimates of the continuous jet intrinsic speed \citep[][Section~\ref{sec:speed and inclination};]{2025ApJ...986L..35Z}. 
            
            For the remainder of the flaring period, the overall radio flux density was somewhat lower, and was partially optically thin, suggesting some quenching of the continuous jet \citep[Figure~\ref{fig:ejection date lightcurve}; see also Figure~4 of][]{2025ApJ...988..109H}. There was some amount of variability between the sparsely measured flux densities. Since these broad radio light curves were made from observations with instruments that capture much larger spatial scales, in the absence of high-resolution VLBI observations, it is not clear if the variable optically thin emission during this period originated from the ongoing repeated ejection of several discrete transient jets, the formation of internal shocks in the continuous jet, or from the continued evolution of previously ejected diffuse downstream transient ejecta (such as knots 1, 2, 3, and 4). 
    
            On the morning of 2023 October 5 (MJD 60222), \sjfull\ was the faintest it had been at radio frequencies during the outburst, with an optically thin spectrum, suggesting that the core jet was largely quenched with the emission dominated by previously diffuse downstream transient ejecta. \citet{2023ATel16273....1B} reported, based on NICER timing observations, that between 03:00-11:00 UTC on MJD 60222, \sjfull\ underwent a hard-intermediate to soft-intermediate state transition. Within hours of this transition, we observed rapid $\sim$hour time-scale flaring in our second \qty{4.9}{\GHz} EVN observation, consistent with the ejection of one or more transient ejecta \citep[Figure~\ref{fig:rm019 amplitudes}; see also][]{2025ApJ...987L..14C}. At the end of this observation, the integrated flux density was around $\sim$\qty{120}{\mJy}, but $\sim$7 hours later, \sjfull\ was detected by the VLA at \qty{235.8\pm1.1}{\mJy} at \qty{5.25}{\GHz}, suggesting continued rapid radio flaring during this period. Following this, our VLBI monitoring revealed the repeated ejection of knots 5, 6, and 7 over the subsequent three days, while daily radio monitoring showed a gradually declining optically thin light curve. As discussed in Section~\ref{sec:variability discussion}, knot 5 showed rapid $\sim$hour time-scale variability that was not observed in the daily radio light curves. During this period, the continuous jet was completely quenched, although it is not clear if we detected it in our second EVN observation, since we could not tell the difference between multiple transient ejecta or a single transient jet knot and the fixed compact continuous jet.
    
            Later, around 2023 October 10 (MJD 60227), \sjfull\ returned to the hard-intermediate state \citep{2023ATel16276....1Y}, and we observed the re-establishment of the continuous jet, where it was compact ($\sim1-2$\ \unit{\mas}) and faint ($\sim2-7$\ \unit{\mJy}), apart from a short-lived radio flare in an observation on 2023 October 11 (MJD 60228). Figure~\ref{fig:ejection date lightcurve} shows that around the time of this observation there was a gradual dip in the X-ray hardness. It is not clear if this flare corresponded to the launching of transient ejecta, nor if the change in X-ray hardness was due to another short excursion to the soft-intermediate state, although we note that following this flare the core was unquenched. This re-establishment of the core jet would have been difficult to identify in the daily radio light curves alone due to contamination from emission from the previously launched, optically thin, diffuse, downstream transient ejecta. This short-lived radio flare was also not captured in the overall radio light curve. \sjfull\ then underwent a final hard-intermediate to soft-intermediate state transition before it eventually headed towards the soft state, which was accompanied by the brightest radio flare of the outburst. We detected and tracked knot 8, and we inferred its ejection date to be $\sim1$~day after the peak of the radio flare and state transition. Based on this, and the repeated jet ejection that occurred during the previous state transition, we suspect multiple transient jet knots were also launched during this final state transition, that we did not detect due to their short-lived nature. Following this final hard-intermediate to soft-intermediate state transition and the eventual transition into the soft state, the continuous jet was quenched, and the overall radio light curves remained optically thin. We then re-detected and resolved the continuous jet following the soft-to-hard reverse transition, where it was faint and marginally resolved.
        
        \subsubsection{Transient Jet Ejection Signatures}\label{sec:jet ejection signatures}
            Our precise ejection date measurements reveal that throughout its 2023 outburst, \sjfull\ repeatedly launched transient jets as it underwent multiple transitions between the hard-intermediate and soft-intermediate states, with multiple ejecta being launched at each transition. The only exceptions to this pattern were knot 0 and knot 4 (although we have not studied the available dense X-ray spectral and timing observations around the time of knot 4's ejection). We interpreted knot 0 as being the result of internal shocks in the continuous jet, and/or interactions between the continuous jet and the interstellar medium or previously ejected jet material. Given its detection alongside the highly extended continuous jet, we cannot rule out that knot 4 could also have a similar origin to knot 0. Figure~\ref{fig:ejection date lightcurve} also shows that while many of the inferred ejection dates correspond to changes in the X-ray intensity and hardness, there was no clear consistent repetitive signature of transient ejection, with jet knots launched at different X-ray luminosities and hardness ratios. We note that while multiple of the jet knots appeared to be ejected close to or during dips in the X-ray hardness, we do not have precise enough ejection date constraints for all of our ejecta to determine if this was a consistent ejection signature throughout the outburst. \change{We also note that changes in the X-ray intensity and hardness do not fully capture the complex evolution of the underlying accretion flow which can be probed in more detail with X-ray spectral modelling and timing studies, which are beyond the scope of this paper.} It is not clear if the launching of multiple transient ejecta within a single outburst (or around the time of a single sudden change in the X-ray properties) requires there to be a common mechanism or signature for each ejection event.
            
            Several other LMXBs are also thought to have launched multiple transient jets within a single outburst (which have either been directly resolved and/or have been inferred from rapid radio flaring), for example: 4U 1543-47 \citep{2025arXiv250411945Z}; XTE J1752-223 \citep{2013MNRAS.432..931B}; XTE J1859+226 \citep{2002MNRAS.331..765B}; MAXI J1820+070 \citep{2020NatAs...4..697B, 2021MNRAS.505.3393W}; and MAXI J1348-630 \citep{2021MNRAS.504..444C}. In more extreme cases, sources that reach Eddington or super-Eddington luminosities during their outbursts can display episodes of rapid and repeated transient jet ejection, and in some cases, with the core remaining unquenched. Examples include: GRO J1655-40 during its bright 1994 outburst \citep{1995Natur.374..703H, 1995Natur.375..464H, 1995Natur.374..141T, 2000ApJ...540..521H}; GRS 1915+105 during multiple bright state transitions \citep[e.g.][]{1994Natur.371...46M, 1999ApJ...511..398R, 1999MNRAS.304..865F}; V404 Cygni during its 2015 outburst \citep{2017MNRAS.469.3141T, 2019Natur.569..374M, 2019MNRAS.482.2950T, 2023MNRAS.518.1243F}; and SS 433 during periods of flaring \citep{2004AAS...20510401S, 2011MNRAS.417.2401B, 2016MNRAS.461..312J}. At a distance of \distanceBurridge, \sjfull\ reached a peak hard-state bolometric luminosity of $1.4_{-0.6}^{+0.9}\ L_\text{Edd}$, where $L_\text{Edd}$ is the Eddington luminosity for a $10M_\odot$ black hole \citep{2025ApJ...994..243B}. It therefore likely exceeded the Eddington luminosity during the flaring state, particularly during the bright X-ray flare on 2023 September 19 (MJD 60206). As discussed in Section~\ref{sec:variability discussion}, at the peak of the outburst, V404 Cygni went through a period of super-Eddington accretion and rapid jet ejection, with the core remaining unquenched, lasting for $\sim14$ days \citep{2017MNRAS.471.1797M, 2019MNRAS.482.2950T, 2019Natur.569..374M,  2023MNRAS.518.1243F}. Towards the end of the peak of the outburst, following the brightest radio flare, high angular resolution observations showed that a bright extended transient jet knot had been launched, and that the core had finally quenched \citep{2019Natur.569..374M}. This seems similar to our observations of \sjfull, where multiple jet knots were launched during the bright, Eddington/super-Eddington flaring period, with the core not becoming completely quenched until the final hard-intermediate to soft-intermediate state transition, which also produced the brightest radio flare. This might suggest that 2023-2024 outburst of \sjfull\ belongs to this class of extreme Eddington/super-Eddington outbursts that show repeated jet ejection, rather than the more canonical outbursts with a single prominent state transition. We note that in V404 Cygni, the inner rapidly-precessing accretion flow was highly obscured, meaning that tracking the precise state of the system was difficult, and no clear transient jet ejection signatures were identified \citep{2019Natur.569..374M}. There were however, differences in these outbursts. In \sjfull, we did not clearly observe the ejection of a transient jet within an individual observation, and so we cannot say if the core was quenched or unquenched at the moment the transient ejecta were launched. The rate at which transient jet ejecta were launched in V404 Cygni was also much greater than in \sjfull, and the jet axis and inner accretion flow of V404 Cygni showed rapid, large-scale precession, which was not observed in \sjfull. Only long, high-cadence, contemporaneous VLBI and X-ray observations will allow us to understand the nature of these extreme outbursts. 
            
            In many LMXBs, the repeated ejection of transient jets has not been associated with repeated signatures of ejection in the inner accretion flow. In GRS 1915+105, the ejection of transient jets are thought to be related to soft X-ray dips \citep{2003ApJ...597.1023V}, possibly due to the repeated ejection of the corona. There have been suggestions that in GRS 1915+105, the corona and the jet are the same physical component, with the corona morphing into the jet \citep{2022NatAs...6..577M, 2022MNRAS.513.4196G}. The suggestion that the corona and the jet are the same physical component date back to \citet{2005ApJ...635.1203M}. Spectral modelling of observations of MAXI J1348-630, MAXI J1535-571, and MAXI J1820+070, also show a clear connection between the evolution of the corona and the jet, finding that the modelled distance of the corona from the black hole increases around the time that the continuous jet is quenched and transient jets are launched \citep{2025ApJ...994...54D}. Although we note that \citet{2019Natur.565..198K} studied X-ray reverberation lags in MAXI J1820+070 and instead found that the corona contracted as the source softened. Observations of \sjfull\ in the flaring period showed rapid evolution of the corona and the disk, with periods of flaring corresponding to the suppression (or ejection) of the corona as the disk extended inwards, enhancing the soft X-rays \citep{2025ApJ...986....3L}. The ejection of the corona has been suggested as the origin of transient jet ejecta in several systems \citep[][]{2003ApJ...595.1032R, 2022ApJ...930...18W}. If the ejection of transient jets is the result of the ejection of the corona, then the corona would have to reform rapidly (within hours) to explain the repeated launching of ejecta seen in \sjfull. This is consistent with observations from the X-ray polarimeter observations which showed that the X-ray polarisation of \sjfull\ was stable over the full state-transition \citep{2024ApJ...968...76I}, which implies that if the corona is linked to transient jet ejection then any drastic changes in the corona (due to jet ejection) must be short-lived.
        
            The association between the ejection of knots 1, 2, and 3, and the bright X-ray flare on 2023 September 19 (MJD 60206) is one of the best connections to date between changes in the inner accretion flow and transient jet ejection in an LMXB. Recently, \citet{2026A&A...706A.208J} analysed a series of HXMT observations during at the peak of this flare, revealing the presence of a type-B QPO for the entire duration of our inferred ejection window. The appearance of type-B QPOs is thought to be marker of the transition from the hard-intermediate state to the soft-intermediate state, and has been shown to appear close to the inferred ejection dates of transient ejecta in a number of LMXBs \citep{2009MNRAS.396.1370F, 2012MNRAS.421..468M, 2019ApJ...883..198R, 2020ApJ...891L..29H, 2021MNRAS.505.3393W}. However, \citet{2026A&A...706A.208J} also showed that the type-B QPO was also present alongside a type-C QPO before and after the bright X-ray flare, and that the apparent switch from type-C to type-B QPOs was due to a suppression of the overall rms and the dominant type-C QPO, revealing the underlying type-B QPO. This new picture is supported by analysis of the disk/jet coupling in MAXI~J1348-630 by \citet{2026A&A...707A.151C}, who found a tentative connection between a drop in the overall X-ray rms and the ejection of transient jets, and did not find any evidence of the appearance of a type-B QPO close to their inferred ejection dates. Although we note that their ejection date uncertainties were between $2-3$ days and their X-ray coverage was far sparser than for \sjfull, meaning short-lived ejection signatures could have been missed. In MAXI~J1820+070, precise measurements of the ejection dates of two transient jets showed that a mildly relativistic jet knot was launched as the type-C QPO faded, just prior to the switch to a type-B QPO, with a second highly relativistic jet knot launched after the disappearance of the type-B QPO \citep{2020ApJ...891L..29H, 2021MNRAS.505.3393W}. It's not clear yet if the type-B QPO in MAXI~J1820+070 was present alongside the type-C QPO, with it being revealed as the type-C QPO and overall rms variability was suppressed, as in \sjfull\ \citep{2026A&A...706A.208J}, or if there was a sudden switch between the two distinct variability modes. Other associations of transient jet ejection with particular signatures of ejection (such as the appearance of type-B QPOs) have often only been constrained to within hours or days due to a combination of a lack of dense contemporaneous X-ray monitoring and precise ejection dates \citep[][]{2012MNRAS.421..468M, 2019ApJ...883..198R, 2021MNRAS.504..444C, 2024MNRAS.533.4188C}. 

            
            Only with precise ejection dates from high angular resolution observations combined with dense X-ray monitoring can we begin to probe the precise sequence of events between particular changes in the inner accretion flow and the ejection of transient jets. While we saw repeated transient jet launching at each hard-intermediate to soft-intermediate state transition, we did not see any specific signatures of repeated jet ejection in the X-ray hardness and intensity. There was also no clear repetitive signature of the individual ejection of knots 1, 2, and 3 in the dense timing observations surrounding and during the bright X-ray flare. Unfortunately, timing observations surrounding the ejection of knots 5, 6, 7, and 8 later in the outburst were sparse due to several instruments becoming sun-constrained, making it difficult to search for signatures of jet ejection in X-ray timing observations.
    
            The recent suggestion of a relationship between large-scale jet precession and the launch speeds of transient jet knots \citep{2025NatAs.tmp..198F} implies that differing properties of transient jets, and in particular their intrinsic speeds, may be closely related to the configuration of the inner accretion flow and the jet launching region. \change{As we discussed in Sections~\ref{sec:position angle and precession} and \ref{sec:speed and inclination}, the jet knots were likely launched along a very similar inclination angle, and therefore the observed differences in the jet properties, in particular their intrinsic speeds, can not be explained by projection or inclination effects.} This large sample of transient jets from a single source allows us to look for potential links between particular X-ray properties and the varying intrinsic properties of the jets launched by \sjfull. \change{However, we found no evidence of a clear connection between the speeds, sizes, and expansion rates of the individual transient jets and the available X-ray intensity and hardness data. We were primarily limited in our ability to search for evidence of this connection due to the cadence and sensitivity of the available X-ray monitoring. We were also limited by the large uncertainties on our ejection date constraints for some of the jet knots (particular knots 3, 5, 6, and 7), which meant that we could not associate these ejection events with precise X-ray hardness or intensity measurements, or with precise changes in these properties.} 
            
            One interesting comparison between the jet launching events that we have identified is the difference between the ejection of knots 1, 2, and 3, and the other ejecta later in the outburst. Knots 1, 2, and 3 were launched during the brightest X-ray flare of the outburst, where there was a drastic increase in the soft X-ray intensity and a drop in the X-ray hardness. While many of the other ejecta were launched around the time of a drop in the X-ray hardness, none of them were launched contemporaneously with a comparable increase in the soft X-ray intensity. This flare, during which \sjfull\ may have briefly transitioned into the soft-intermediate state \citep{2026A&A...706A.208J}, was also much more short-lived than the later transitions from the hard-intermediate state to the soft-intermediate state. Compared to knots 1, 2, and 3, the ejecta launched later during the two prominent hard-intermediate to soft-intermediate state transitions generally had larger flux densities in our VLBI observations, and were accompanied by much brighter radio flares (as detected in the sparse radio monitoring). The accretion/ejection coupling during the flaring period (which may have been a hard-intermediate state or a very high state; see Section~\ref{sec:Swift j1727 introduction}) may have been different than during the two prominent state-transitions later in the outburst. Careful and detailed analysis and modelling of the spectral and timing features in the available X-ray observations surrounding our inferred ejection dates are required to further investigate the nature of the complex accretion/ejection coupling in \sjfull. 

\section{Conclusions}
    We have presented the results from our comprehensive VLBI monitoring campaign of \sjfull\ in its 2023-2024 outburst, which consisted of 16 observations from the VLBA, the LBA, and the EVN. Our observations spanned the bright hard/hard-intermediate state, the flaring state, throughout and between the two prominent hard-intermediate to soft-intermediate state transitions, and the eventual soft-to-hard reverse transition. We tracked the evolution of the continuous jet through these states, finding that it faded and became less extended as \sjfull\ moved through the hard-intermediate and flaring states, and that it quenched and reformed several times following episodes of radio flaring and transient jet launching. We also found evidence of a luminosity-dependent core shift. 
    
    Using time-dependent visibility model fitting, we precisely measured the motions, sizes, and flux density variability of nine individual approaching jet knots, all moving to the south of the core at a relatively consistent position angle (to within $\sim2.5^\circ$ of each other). We did not detect any receding counterparts to these jet knots. We also found evidence of more transient jet ejecta that we were unable to model, due to their rapid variability. We argue that at least one of the jet knots was likely a short-lived internal shock in an underlying highly-extended continuous jet, while the others were discrete ejecta repeatedly launched during several hard-intermediate to soft-intermediate state transitions. During each of these transitions, $\sim$daily cadence radio monitoring showed bright radio flares that decayed over the course of several days, however our high angular resolution observations revealed that each of these flares corresponded to the repeated and rapid ejection of multiple transient ejecta over the span of multiple days. These transient jets were only detected for a short time on VLBI scales, and so without dense VLBI observations analysed with time-dependent visibility modelling, we would not have been able to uncover the rapid ejection of transient jets during this period. We also suggest that more transient ejecta were likely launched during the outburst that we could not detect due to gaps in our monitoring. While the ejection of the transient jet knots that we detected generally occurred either during or immediately following the several repeated hard-intermediate to soft-intermediate state transitions, we did not find a single consistent repeated signature of jet ejection in \sjfull\ \change{in the available X-ray hardness and intensity data. We note that a lack of dense X-ray timing observations for periods of the outburst limited our ability to search for jet ejection signatures in the X-ray timing properties}. For the first time, we fit the intra-observation light curves of the ejecta with non-parametric piecewise models, revealing that some of these ejecta showed rapid $\sim$hour time-scale variability that was not captured in the overall radio light curves. The jet knots also did not expand and evolve with a constant opening angle, suggesting that simple models of transient jet evolution where they expand adiabatically and evolve steadily in flux density are not adequate to explain their properties and evolution. The transient jet knots were all more laterally extended than the extended continuous jet, suggesting that they were less confined. Using a novel approach, we have also used the measured proper motions of the approaching transient jet knots to constrain their intrinsic speeds without having to precisely constrain their inclination, finding that \sjfull\ repeatedly launched both mildly relativistic ($\beta\Gamma<1$) and highly relativistic ($\beta\Gamma>2$) ejecta. From the proper motions of these ejecta, we constrain the 50th, 84th, and 99th percentile upper limits of the jet axis inclination to be $40^\circ$, $50^circ$, and $66^\circ$, respectively. These observations suggests that fixed parameters for an individual LMXB, such as black-hole mass, black-hole spin, and spin-orbit misalignment, do not uniquely determine the varying properties of transient jet knots, including their speeds and Lorentz factors.
    
    Without time-dependent visibility modelling, we would not have been able to determine most of the transient jet properties that we have precisely measured, including their motions, ejection dates, and intra-observation light curves, significantly reducing the value of these high-resolution observations. Despite this, there are still many unanswered questions, particularly, about the mechanisms responsible for the launching of transient jets and the connection between their varied properties and the configuration of the inner accretion flow. Future dense, long-duration, high-angular resolution observations around the peak of the outburst are essential to precisely track the properties and evolution of transient jets. With time-dependent visibility model fitting, we no longer need to track the motions of transient jet knots between daily observations, since we can precisely measure their motions and flux density evolution within a single observation. This will allow us to focus on taking longer duration observations at the peak of the outburst, as close as possible to the state-transition, to capture the transient jets knots as soon as possible after they are launched. This will require rapid X-ray triggers to identify these critical periods in real-time, and dense contemporaneous X-ray monitoring to allow us to identify ejection signatures to determine the precise causal connection between the changes in the inner accretion flow and the launching of transient jets.

    The increased sensitivity and greater angular resolution of future VLBI instruments, like the next-generation Event Horizon Telescope (ngEHT) and SKA-VLBI, will allow for more detailed studies of transient ejecta close to their launch dates. However, the intra-observation motions and variability of these jets will be even more significant in these observations, and we will require techniques like time-dependent visibility modelling to precisely track and reconstruct their real-time evolution.

\section*{Acknowledgements}
    We respectfully acknowledge the significant contributions made to this longstanding collaboration by Tomaso Belloni, who sadly passed away during our observing campaign. His insights and wealth of knowledge are sorely missed by his colleagues.
    
    The National Radio Astronomy Observatory is a facility of the National Science Foundation operated under cooperative agreement by Associated Universities, Inc. This work made use of the Swinburne University of Technology software correlator, developed as part of the Australian Major National Research Facilities Programme and operated under licence. The Long Baseline Array is part of the Australia Telescope National Facility (\url{https://ror.org/05qajvd42}) which is funded by the Australian Government for operation as a National Facility managed by CSIRO. This work was supported by resources provided by the Pawsey Supercomputing Research Centre with funding from the Australian Government and the Government of Western Australia. \change{The European VLBI Network is a joint facility of independent European, African, Asian, and North American radio astronomy institutes. Scientific results from data presented in this publication are derived from the following EVN project code(s): RM018, RM019 \citep{RM019}. e-VLBI research infrastructure in Europe is supported by the European Union’s Seventh Framework Programme (FP7/2007-2013) under grant agreement number RI-261525 NEXPReS.} This research has made use of the MAXI data provided by RIKEN, JAXA and the MAXI team. This work made use of the Warkworth 30m telescope as part of the LBA \citep{2015PASA...32...17W}. From 2023 July 1, operation of Warkworth was transferred from Auckland University of Technology (AUT) to Space Operations New Zealand Ltd, who continue to make the facilities available for VLBI out of goodwill. 
    
    CMW acknowledges financial support from the Forrest Research Foundation Scholarship, the Jean-Pierre Macquart Scholarship, and the Australian Government Research Training Program Scholarship. TDR is an INAF research fellow. PA is supported by the WISE fellowship program, which is financed by NWO. AJT acknowledges that this research was undertaken thanks to funding from the Canada Research Chairs Program and the support of the Natural Sciences and Engineering Research Council of Canada (NSERC; funding reference number RGPIN--2024--04458). RF and SM are supported by a European Research Council (ERC) Synergy Grant "BlackHolistic" grant No. 10107164. DMR is supported by Tamkeen under the NYU Abu Dhabi Research Institute grant CASS. VT acknowledges support from the Romanian Ministry of Research, Innovation and Digitalization through the Romanian National Core Program LAPLAS VII – contract no. 30N/2023.
    
    The authors wish to recognise and acknowledge the very significant cultural role and reverence that the summit of Maunakea has always had within the indigenous Hawaiian community. We are most fortunate to have the opportunity to conduct observations from this mountain. We also wish to acknowledge the Gomeroi, Gamilaroi, and Wiradjuri people as the traditional custodians of the LBA observatory sites. 

\section*{Data Availability}
The data used in this work are available via reasonable request to the corresponding author.

\bibliographystyle{aasjournal}
\bibliography{references}

\end{document}